\documentclass[%
reprint,
superscriptaddress,
 amsmath,amssymb,
 aps,
 prd,
 numerical,
 longbibliography,
]{revtex4-2}

\usepackage{graphicx}
\usepackage{dcolumn}
\usepackage{bm}

\usepackage[varg]{txfonts}
\usepackage[T1]{fontenc}
\usepackage[utf8]{inputenc}   
\DeclareUnicodeCharacter{2243}{\ensuremath{\simeq}}
\usepackage[colorlinks=true, allcolors=blue]{hyperref}
\usepackage{xcolor}
\usepackage{float}
\usepackage{aas_macros_prd}
\usepackage[switch]{lineno}

\usepackage{orcidlink}

\defcitealias{Moretti_2015}{M15}
\defcitealias{Richard_2021}{R21}
\defcitealias{Paraficz_2016}{P16}
\defcitealias{Cha_2025}{C25}

\begin{document}


\title{Baryonic mass budgets in the central regions of the Bullet Cluster and
    their consistency with strong lensing in MOND} 


\author{Dong Zhang\orcidlink{0000-0002-9011-8875}}
\email{zhang\_dong\_96@163.com}
\affiliation{Helmholtz-Institut f\"ur Strahlen-und Kernphysik (HISKP), University of Bonn, Nussallee 14--16, D-53115 Bonn, Germany}

\author{Hosein Haghi\orcidlink{0000-0002-9058-9677}}
\email{haghi@iasbs.ac.ir}
\affiliation{Department of Physics, Institute for Advanced Studies in Basic Sciences (IASBS), Zanjan 11365-9161, Iran}
\affiliation{Helmholtz-Institut f\"ur Strahlen-und Kernphysik (HISKP), University of Bonn, Nussallee 14--16, D-53115 Bonn, Germany}
\affiliation{School of Astronomy, Institute for Research in Fundamental Sciences (IPM), Tehran 19395-5531, Iran}

\author{Elena Asencio\orcidlink{0000-0002-3951-8718}}
\affiliation{Helmholtz-Institut f\"ur Strahlen-und Kernphysik (HISKP), University of Bonn, Nussallee 14--16, D-53115 Bonn, Germany}

\author{Indranil Banik\orcidlink{0000-0002-4123-7325}}
\email{indranil.banik@port.ac.uk}
\affiliation{Institute of Cosmology and Gravitation, University of Portsmouth, Dennis Sciama Building, Burnaby Road, Portsmouth PO1 3FX, UK}

\author{Akram Hasani Zonoozi\orcidlink{0000-0002-0322-9957}}
\affiliation{Helmholtz-Institut f\"ur Strahlen-und Kernphysik (HISKP), University of Bonn, Nussallee 14--16, D-53115 Bonn, Germany}
\affiliation{Department of Physics, Institute for Advanced Studies in Basic Sciences (IASBS), Zanjan 11365-9161, Iran}

\author{Sangjun Cha\orcidlink{0000-0001-7148-6915}}
\email{sang6199@yonsei.ac.kr}
\affiliation{Department of Astronomy, Yonsei University, 50 Yonsei-ro, Seoul 03722, Republic of Korea}

\author{Boseong Young Cho\orcidlink{0009-0009-4273-6132}}
\email{boseongcho@yonsei.ac.kr}
\affiliation{Department of Astronomy, Yonsei University, 50 Yonsei-ro, Seoul 03722, Republic of Korea}

\author{Hyungjin Joo\orcidlink{0000-0001-9139-5455}}
\email{gudwls4478@yonsei.ac.kr}
\affiliation{Department of Astronomy, Yonsei University, 50 Yonsei-ro, Seoul 03722, Republic of Korea}

\author{Pavel Kroupa\orcidlink{0000-0002-7301-3377}}
\affiliation{Helmholtz-Institut f\"ur Strahlen-und Kernphysik (HISKP), University of Bonn, Nussallee 14--16, D-53115 Bonn, Germany}
\affiliation{Astronomical Institute, Faculty of Mathematics and Physics, Charles University, V Hole\v{s}ovi\v{c}k\'ach 2, CZ-180 00 Praha 8, Czech Republic}

\author{Anastasiia Lazutkina\orcidlink{0009-0007-3964-1525}}  
\affiliation{Interdisciplinary Centre for Science and Technology Studies (IZWT), University of Wuppertal, Gau{\ss}stra{\ss}e 20, Wuppertal 42103, Germany}

\author{Eda Gjergo\orcidlink{0000-0002-7440-1080}}
\affiliation{School of Astronomy and Space Science, Nanjing University, Nanjing, 210093, Jiangsu, People's Republic of China}
\affiliation{Key Laboratory of Modern Astronomy and Astrophysics, Nanjing University, Ministry of Education, Nanjing, 210093, Jiangsu, People's Republic of China}

\date{\today}




\begin{abstract} 
Strong \emph J\emph W\emph S\emph T lensing observations of the Bullet Cluster have traditionally been regarded as definitive evidence for the existence of dark matter, thereby posing a major challenge to the Milgromian dynamics (MOND) framework. The offset between the lensing mass and the X-ray gas centroids indicates the presence of a substantial amount of unseen mass in the vicinity of the brightest cluster galaxy (BCG). However, the high metallicities observed in the intracluster X-ray gas and in the massive early-type member galaxies suggest a past epoch dominated by massive stars, which have since evolved into stellar remnants. Their mass contribution is naturally incorporated in the integrated galaxy-wide initial mass function (IGIMF) theory, which predicts a significantly higher baryonic mass in early-type galaxies than the canonical IMF. In this work, we re-estimate the baryonic masses of the three BCG-centred core regions of the Bullet Cluster using recent JWST photometry and compare them with MOND strong-lensing masses. The IGIMF masses are derived for constant- and for (self-) enriched-metallicity stellar population synthesis models. These constitute, respectively, the lower and the upper limits of the IGIMF estimated mass, though the latter is the more realistic self-consistent scenario. Our results show that the MOND strong lensing masses of all cores fall between the IGIMF masses derived from the constant- and the enriched-metallicity stellar models. 
These results suggest that the baryonic mass budget is consistent with MOND requirements from strong-lensing observations in the core regions of the Bullet Cluster. However, the physical viability of this scenario also depends on the spatial distribution and dynamical behavior of the remnant population, which remain to be established. 
Independently of the validity of MOND, there is now an emerging perspective regarding dark matter models: to account for the observations, they seem to require less dark matter than previously anticipated.



\end{abstract}

\maketitle


\section{Introduction}
\label{introduction}

The Bullet Cluster (1E 0657--558) is composed of two colliding sub-clusters at $z=0.296$ \citep{Tucker_1998}. It has been commonly regarded as ``a direct empirical proof of the existence of dark matter'' \citep{Clowe_2006}. This claim arises from the X-ray and weak-lensing observations of the cluster, which show an offset between its hot bright gas component at the centre of the system and the faint regions at the edges, where weak lensing observations indicate most of the cluster's mass to be.

In the Lambda cold dark matter ($\Lambda$CDM) paradigm \citep*{Efstathiou_1990, Ostriker_1995}, which is also known as the standard model of cosmology, dark matter interacts with ordinary matter primarily through gravity, while all other interactions are assumed to be negligible.
Therefore, in this model, the mass distribution and the offset between the different components of the Bullet Cluster are explained by the fact that the gas interacted electromagnetically with itself during the collision, causing it to slow down with respect to the dark matter component, which only interacts gravitationally. In this regard, the Bullet Cluster has been presented not only as proof that dark matter exists, but also as a challenge to theories that explain the dynamics of galaxies without it.

Milgromian dynamics \citep[MOND;][]{Milgrom_1983,Bekenstein_Milgrom_1984} is currently one leading model among these theories. In MOND, gravity experiences a boost with respect to Newton's law when the gravitational acceleration falls below $ a_0 = 1.2 \times 10^{-10}~\textrm{m/s}^2$ \citep*{Begeman_1991, Gentile_2011, McGaugh_2016_RAR}. MOND has proven to be very successful in explaining several phenomena on galactic scales, even more so than $\Lambda$CDM \citep[for reviews, see][]{Famaey_McGaugh_2012, Banik_Zhao_2022, Kroupa_2023}. However, it has also been found that the MONDian boost to Newtonian gravity is insufficient to fully explain the dynamics of galaxy clusters, as the predicted gravity for these systems using their baryonic mass alone is still lower than the observed gravity \citep{The_1988, Gerbal_1992, Sanders_1999, Sanders_2003, Pointecouteau_2005, Natarajan_2008, Angus_2008, Ettori_2019, Famaey_2024, Kelleher_2024}.

It is sometimes assumed that the discovery of dark matter, either through laboratory experiments or through observations of galaxy clusters, would rule out MOND. However, the existence of dark matter particles does not in itself violate the basic principles of MOND. For instance, \citet{Angus_2009} proposed a hybrid model named ``neutrino-Hot-Dark-Matter'' ($\nu$HDM) in which MOND is still the gravitational law determining the dynamics of galaxies and galaxy clusters, but the source of this gravity on galaxy cluster scales also includes the gravitational contribution of weakly-interacting particles, the $11\textrm{eV}/c^2$ rest energy hot sterile neutrinos ($\nu$). Although tests on the viability of $\nu$HDM as a realistic cosmological model are still ongoing \citep{Katz_2013, Wittenburg_2023, Samaras_2025,Russell_2026MNRAS}, this could be a promising model since it combines the advantages of MOND and $\Lambda$CDM, making it possible to explain most of the observed astrophysical and cosmological phenomena \citep{Angus_2009, Angus_2010, Haslbauer_2020}. $\nu$HDM is consistent with the exclusion of cold or warm dark matter particles based on the Chandrasekhar dynamical friction test \citep{Haslbauer_2024_MCs,Hernandez_2025} and the distinctly asymmetric tidal tails of open star clusters \citep{Kroupa_2024_open}, which are difficult to understand in linear theories of gravity. $\nu$HDM also better accounts for the observed signs of tidal disturbance in dwarf galaxies in the Fornax Cluster and the lack of low surface brightness dwarfs towards its centre, which are difficult to reconcile with their expected stability to cluster tides in the $\Lambda$CDM framework\citep{Asencio_2022}.

Even if not sufficient to rule out MOND, the Bullet Cluster is often interpreted as indicating that the missing mass problem requires a non-interacting (or weakly-interacting) component, based on the observed separation between the X-ray emitting gas and the weak-lensing peaks. However, alternative interpretations have been discussed in the literature. For example, \citet{Jan_2025} pointed out that several assumptions underlying the standard analysis may introduce systematic uncertainties. 
In particular, given that the sound crossing time is larger than the time since the collision of the two sub-clusters, hydrostatic equilibrium of the hot X-ray emitting gas may not be valid.
The X-ray mass of the Bullet Cluster, obtained under the incorrect assumption that the cluster is in hydrostatic equilibrium, is likely an overestimation of the hot gas mass \citep{Markevitch_2002}. A lower gas mass would reduce the apparent discrepancy with the lensing-derived mass distribution. 
Additionally, \citet{Jan_2025} noted that, in the original work of \citet{Clowe_2006}, there are actually four mass peaks for the X-ray gas, with similar mass in each peak. Two of the peaks correspond to the well-known central peaks, but the other two are located at the brightest galaxy of each sub-cluster, within the regions in which the weak-lensing indicates that most of the mass is contained. This suggests that the degree of separation between the gas and the total mass distribution may be less clear-cut than often assumed.

The weak-lensing peaks of the Bullet Cluster around the BCGs are traditionally attributed to dark matter, but they also coincide with the peaks in the galaxy number density. One could argue that this is merely a coincidence due to both the dark matter and the galaxies being largely collisionless. This raises the important question of how much mass each galaxy has, and thus how much (if any) extra contribution must come from a non-baryonic component. The stellar mass $M_\star$ must be estimated from the photometry, but we also need to account for galaxies that fall below the detection limit given the Gpc distance to the Bullet Cluster. Additionally, the Bullet Cluster contains hot gas (visible in X-rays) and intracluster light (ICL), which is assumed to originate mostly from stellar stripping of galaxies by each other and the overall cluster potential. We revisit the contribution of each component to the baryonic mass of the Bullet Cluster, as this is crucial for evaluating the mass budget underlying the observed lensing signal.

In order to account for the presently non-observable baryonic mass coming from stellar remnants (especially neutron stars and black holes), it is important to consider the constraints imposed by the metallicity detected in galaxy clusters. \citet*{Blackwell_2022} found that the stellar populations of the massive galaxy clusters in their sample were clearly insufficient to account for the observed intracluster metallicity. This suggests that the high metallicity was generated by an earlier population of very massive stars. The remnants of this early population will therefore constitute an important contribution to the total mass budget of galaxy clusters.

The integrated galaxy-wide initial mass function \citep[IGIMF;][]{Kroupa_2003_IGIMF} theory allows an inference of the total stellar mass-to-light ($M_\star/L$) ratio of galaxies, where the mass includes both living stars and stellar remnants. This theory extends the traditional stellar initial mass function \citep[IMF;][]{Kroupa_2001} to describe the mass distribution of stellar populations throughout a galaxy. The IGIMF theory proposes that massive galaxies contain a higher proportion of massive stars at early times \citep{Jerabkova_2018, Yan_2021, Zonoozi_2025_SPS, Kroupa_2026_IMF}. These stars die quickly and leave behind dark remnants—white dwarfs, neutron stars, or black holes—while also producing a present-day bottom-heavy galaxy-wide IMF (gwIMF). Both effects act to increase the present-day $M_{\star}/L$. This allows us to reproduce the high metallicities observed in present-day systems \citep{Loewenstein_2013}. It is important to emphasize that the IGIMF theory has been gauged on a large amount of observational data, spanning from young star clusters through the Galactic field population to major starbursts at high redshift, thereby constituting a well-established framework for calculating stellar populations in different environments. The IGIMF theory has been developed entirely independently of MOND.

Since previous studies of the baryonic mass content of the Bullet Cluster have not explicitly accounted for the potential contribution of stellar remnants, it is relevant to revisit the ``missing mass problem'' in this system within the IGIMF framework.
The ``missing mass'' needed to match the lensing observations is significantly smaller in MOND than in $\Lambda$CDM. This raises the possibility that the additional baryonic mass predicted by the IGIMF theory, combined with the MOND enhancement of gravity, may help to alleviate the discrepancy between the predicted and observed gravitational field, e.g., from lensing observations.

If a significant fraction of the mass in the Bullet Cluster can be accounted for by baryonic matter alone, this would have implications for the interpretation of the system within different theoretical frameworks. In particular, it would reduce the need for an additional collisionless component in MOND-based descriptions. In previous studies \citep{Kraljic_2015}, $\Lambda$CDM simulations have shown that the extreme properties of mass, infall velocity and small sub-cluster distance of the Bullet Cluster are fairly difficult to reproduce \citep*[$2.78\sigma$ tension with $\Lambda$CDM according to][]{Asencio_2021}. Some cosmological simulations that assume MOND as the gravitational law already indicate that structures with very high mass and infall velocity are easier to reproduce in MOND \citep{Katz_2013, Wittenburg_2023, Samaras_2025}.

In this work, we construct stellar population synthesis (SPS) models of galaxies under the IGIMF framework, adopting both constant-metallicity and enriched-metallicity evolutionary assumptions. We derive the galaxy $M_\star/L$ ratio (including stars and their remnants) in different bands. Using these ratios, we re-estimate the total baryonic mass in the regions surrounding the three BCGs of the Bullet Cluster. For this purpose, we employ the member-galaxy catalog of the Bullet Cluster constructed from publicly available JWST/NIRCam imaging obtained in GO-4598 (PI: Maru\v{s}a Brada\v{c}), as used in \citet{Cha_2025} (hereafter \citetalias{Cha_2025}), together with the X-ray gas distribution model derived from observations by \citet{Paraficz_2016}. We then compare the total baryonic masses within the corresponding radii with the MOND strong lensing masses, i.e., the masses required by MOND within the same projected radii to explain the observed strong lensing, as an assessment of MOND in the central regions of the Bullet Cluster.

The paper is organized as follows: In Section~\ref{Relativistic}, we describe our approach to relativistic MOND in the weak deflection limit. Section~\ref{IGIMF} describes the novel application of the IGIMF theory to galaxy clusters. In Section~\ref{IGIMF-Bullet}, we apply the IGIMF theory to estimate the baryonic masses in the core regions around the three BCGs in the Bullet Cluster. We then discuss our results in Section~\ref{Discussion} and conclude in Section~\ref{Conclusions}. The Hubble constant adopted throughout this paper is $H_{0} = 70~{\rm km~s^{-1}~Mpc^{-1}}$. We adopt the cosmological density parameters $\Omega_{\rm m} = 0.3$ and $\Omega_{\Lambda} = 0.7$, thereby assuming a flat cosmology.

\section{Bullet Cluster lensing with relativistic MOND theories} 
\label{Relativistic}

Since the Bullet Cluster is a collision between two sub-clusters, the lack of virial equilibrium makes a traditional dynamical mass inaccurate. We therefore obtain constraints on its gravitational field using instead strong lensing observations \citep{Paraficz_2016, Richard_2021, Cha_2025}. Interpreting these requires a relativistic theory. Traditionally, lensing data is interpreted in general relativity \citep[GR;][]{Einstein_1915}, the simplest relativistic formulation incorporating Newtonian gravity. GR predicts a particular relation between the gravitational field $\bm{g}$ felt by non-relativistic particles and the deflection angle of a photon \citep[equation~43 of][]{Banik_Zhao_2022}:
\begin{eqnarray}
    \Delta \widehat{\bm{n}} ~=~ \int \frac{2 \bm{g}}{c^2} \, dl \, ,
    \label{Light_deflection_angle}
\end{eqnarray}
where $\widehat{\bm{n}}$ is the direction of travel of the photon, $c$ is the speed of light, and $dl$ is a unit of length along its unperturbed trajectory. As discussed in their section~2.6, this relation is preserved in viable relativistic theories of MOND, for instance in Tensor-Vector-Scalar gravity \citep[TeVeS;][]{Bekenstein_2004} and its more recent modifications designed to address the gravitational wave event GW170817, which showed that gravitational waves travel at $c$ \citep{LIGO_Virgo_2017, Skordis_2019}. Assuming Equation~(\ref{Light_deflection_angle}), it was shown that it is possible to explain the offset between the X-ray gas and the regions containing the galaxies, but there still seems to be a ``missing'' component in the total gravitational contribution \citep{Angus_2007}. A key aspect of these analyses is the estimated baryonic mass content of the Bullet Cluster, which we revisit next. Several other relativistic MOND models have been proposed in the last few years \citep{Bernal_2011, Skordis_2021, Monjo_2023, Monjo_2025a, Monjo_2025b}, but these have so far not been tested against the Bullet Cluster observations.

Since Equation~(\ref{Light_deflection_angle}) is expected to be valid in MOND, we can apply it to the specific case of the Bullet Cluster to work out its strong lensing mass. In the context of GR, \citetalias{Cha_2025} derived a strong lensing mass of $M_{\rm lens}^{\rm Newton} = (0.87 \pm 0.13) \times 10^{14} \, M_\odot$ within 250~kpc for the central region of the subclump (also known as the subcluster) of the Bullet Cluster. We can define a suitably weighted average gravity $\overline{g}$ such that the deflection angle
\begin{eqnarray}
    \Delta \theta ~\equiv~ \frac{4b\overline{g}}{c^2} \, ,
\end{eqnarray}
where the impact parameter $b = 250$~kpc. In GR, Equation~(\ref{Light_deflection_angle}) implies that $\overline{g} = GM/b^2$, the Newtonian gravity at closest approach if the lens is treated as a spherical mass concentrated entirely within radius $r \leq b$. Since the Newtonian lensing mass is derived using GR, we must have that
\begin{eqnarray}
    \overline{g}_{\rm obs} ~=~ \frac{GM_{\rm lens}^{\rm Newton}}{b^2} \, .
\end{eqnarray}
This highlights that with an inverse square gravity law at $r \geq b$, there is a simple relation between the gravity at closest approach and the total deflection angle $\Delta \theta$, even though this depends on the gravity at $r \geq b$. This dependence on larger radii will become important below.

In the weak deflection limit of GR, the linear relationship between the gravitational field and the mass distribution implies that the deflection angle depends only on $M_{\rm p}$, the mass within projected radius $r \leq b$, provided the surface density distribution is axisymmetric. This is because dispersing the mass along the line of sight would have no impact on the deflection angle, provided we can make the ``thin lens approximation'' that the mass is concentrated in a region much smaller than the distance from Earth to the lens (Equation~\ref{Light_deflection_angle}). As a result, $\overline{g}_{\rm obs}$ and thus $M_{\rm lens}^{\rm Newton}$ directly tell us $M_{\rm p}$, which we can then compare to the estimated baryonic mass within the same aperture.

This simplification is not valid in MOND because the gravitational field is non-linear in the mass distribution \citep{Banik_Zhao_2022}. If the mass is dispersed along the line of sight, the gravitational field will generally become weaker, pushing the system further into the MOND regime. This makes it important to have a 3D model for the mass distribution, not merely a 2D model for the surface density distribution. We assume that the mass distribution follows a Plummer profile \citep{Plummer_1911}.
\begin{eqnarray}
    M \left( < r \right) ~=~ \frac{Mr^3}{\left( r^2 + a^2 \right)^{3/2}} \, ,
    \label{Plummer_profile_3D}
\end{eqnarray}
where $M$ is the total mass, $M \left( < r \right)$ is the mass within 3D radius $r$, and $a$ is the Plummer scale radius. This leads to the enclosed projected mass behaving as follows:
\begin{eqnarray}
    M_{\rm p} \left( < r_{\rm p} \right) ~=~ \frac{Mr_{\rm p}^2}{r_{\rm p}^2 + a^2} \, .
    \label{Plummer_profile_2D}
\end{eqnarray}

In MOND, we usually need a smaller mass to achieve the same lensing effect \citep{Ferreras_2008, Ferreras_2012, Chang_2013}. To simultaneously solve for both $M$ and $a$, we use $\overline{g}_{\rm obs}$ measurements at two projected radii, which we denote $b_{\rm in}$ and $b_{\rm out} > b_{\rm in}$. Starting with some initial guess for $\left( M, a \right)$, we use the Newton-Raphson procedure to match both observed values of $\overline{g}_{\rm obs}$, as described below.

We can apply Equation~(\ref{Light_deflection_angle}) in both GR and MOND. Writing it as an integral in the 3D radius $r$, we get that
\begin{eqnarray}
    \overline{g} ~=~ \int_b^\infty \frac{g \left( r \right) \, dr}{\sqrt{r^2 - b^2}} \, .
    \label{g_bar}
\end{eqnarray}
The gravity $g$ at radius $r$ is enhanced over the Newtonian gravity $g_{\rm _N} \equiv GM \left( < r \right)/r^2$ by the factor $\nu$ according to the simple interpolating function \citep{Famaey_2005}:
\begin{eqnarray}
    g ~=~ g_{\rm _N} \nu \left( g_{\rm _N} \right) ~=~ \frac{g_{\rm _N}}{2} + \sqrt{ \left( \frac{g_{\rm _N}}{2} \right)^2 + g_{\rm _N} a_0} \, .
    \label{Simple_IF}
\end{eqnarray}
This has been shown to fit the observed RAR in disc galaxies very well for accelerations within $\approx 1$~dex of $a_0 = 1.2 \times 10^{-10}$~m/s\textsuperscript{2} \citep[see figure~23 of][]{Banik_2024_WBT}. Recent studies have pointed out that the rather gradual return to Newtonian behaviour at high accelerations creates serious challenges for this interpolating function on the smaller scale of the Solar System \citep{Hees_2014, Hees_2016, Desmond_2024, Vokrouhlicky_2024}. It is possible that larger scales like galaxy clusters could also require a different interpolating function to that in galaxies. The effect of this on the MONDian mass estimate remains to be investigated in future studies. We note that compared to the difference in scale between the Galaxy and the Solar System, the difference between galaxy and cluster scales is far smaller, justifying our assumption of the same interpolating function in both.

We numerically evaluate the integral in Equation~(\ref{g_bar}) out to $10^4 \, b$ and find the power-law dependence of $g$ at that radius, i.e., we evaluate $\alpha \equiv \mathrm{d} \ln g/\mathrm{d} \ln r$ at $r = 10^4 \, b$ and assume that $g \propto r^\alpha$ at larger radii. Since $r \gg b$, we approximate the denominator at lowest order in $b/r$ as follows:
\begin{eqnarray}
    \frac{1}{\sqrt{r^2 - b^2}} ~\approx~ r^{-1} \left( 1 + \frac{b^2}{2 r^2} \right) \, .
\end{eqnarray}
With these approximations, the integrand in Equation~(\ref{g_bar}) becomes a sum of two power-laws in $r$, allowing us to analytically estimate the contribution from $r \geq 10^4 \, b$.

To avoid a divergent behaviour as $r \to b$, we perform the numerical integral using a linear grid in the variable $\theta$, which we define such that
\begin{eqnarray}
    r ~\equiv~ b \cosh \theta \, .
\end{eqnarray}
Over each interval in $\theta$, we approximate that $g$ is linear in $r$, allowing us to analytically find $\Delta \overline{g}$, the contribution to $\overline{g}$ from the range of $\theta$ under consideration. To maximise the accuracy of our result, we find $g$ and $g' \equiv \mathrm{d}g/\mathrm{d}r$ at both the start, middle, and end of each interval in $\theta$. This leads to three different linear fits to $g \left( r \right)$ and three estimates for $\Delta \overline{g}$. We distinguish these with subscripts $s$, $m$, and $e$ to indicate if the linear fit to $g \left( r \right)$ is based on $g$ and $g'$ at the start, middle, or end, respectively, of the interval in $\theta$. We then combine these estimates using the following weights:
\begin{eqnarray}
    \Delta \overline{g} ~=~ \frac{8 \Delta \overline{g}_{\rm m} - \Delta \overline{g}_{\rm s} - \Delta \overline{g}_{\rm e}}{6} \, .
\end{eqnarray}
This differs from the standard Runge-Kutta weights because we use both $g$ and $g'$. It can be verified that if this procedure is applied to integrate a function, the estimated integral will be correct up to fourth order in the function. The unusual negative weights on $\Delta \overline{g}_{\rm s}$ and $\Delta \overline{g}_{\rm e}$ can be understood by considering the integral of $x^2$ over the range 0 to 1, in which case all three estimates of $\Delta \overline{g}$ underestimate the true value, while  $\Delta \overline{g}_{\rm m}$ does so to a smaller extent.

We apply the above procedure to estimate $\overline{g}_{\rm in}$ and $\overline{g}_{\rm out}$ using impact parameters of $b_{\rm in}$ and $b_{\rm out}$, respectively. Using the Newton-Raphson algorithm, we vary $M$ and $a$ to ensure that the numerically evaluated MONDian $\overline{g} = \overline{g}_{\rm obs}$ for both apertures, indicating that we have matched both observational constraints. Given our result for the Plummer parameters, we then apply Equation~(\ref{Plummer_profile_2D}) to find the predicted $M_{\rm p}^{\rm in}$ and $M_{\rm p}^{\rm out}$, the projected mass within the inner and outer aperture, respectively. For the inner aperture of 100~kpc, we find that the MOND prediction for $M_{\rm p}$ is about $1.6\times$ smaller than $M_{\rm lens}^{\rm Newton}$ even though $\overline{g}_{\rm obs} \approx 7.0 \, a_0$. MOND still predicts a non-negligible enhancement to the GR lensing signal due to the need to integrate contributions to the lensing along the line of sight, which necessarily probes regions with weak gravity. 

We propagate uncertainties in $M_{\rm p}^{\rm in}$ and $M_{\rm p}^{\rm out}$ assuming that these depend linearly on the observed $M_{\rm lens}^{\rm Newton}$ within each aperture. By varying these two observables by $\pm$ their uncertainty and quantifying the impact on $M_{\rm p}^{\rm in}$ and $M_{\rm p}^{\rm out}$, we build up a $2 \times 2$ Jacobian matrix $\mathbf{J}$ relating the observed $M_{\rm lens}^{\rm Newton}$ values to the predicted $M_{\rm p}$ within each aperture.
\begin{eqnarray}
    \mathbf{J} ~\equiv~ \frac{\partial \bm{M}_{\rm p}}{\partial \bm{M}_{\rm lens}^{\rm Newton}} \, .
\end{eqnarray}
If we wish to change $\bm{M}_{\rm p}$ by a certain amount, the inverse matrix $\mathbf{J}^{-1}$ tells us how much we would need to change the underlying $\bm{M}_{\rm lens}^{\rm Newton}$ observables. Since we know their uncertainties, we can then find the $\chi^2$ associated with any shift $\Delta \bm{M}_{\rm p}$ in $\bm{M}_{\rm p}$ space from the $\bm{M}_{\rm p}$ values that arise from using both observed $M_{\rm lens}^{\rm Newton}$ values. In this way, we define a cost matrix $\mathbf{C}_{\rm p}$ in $\bm{M}_{\rm p}$ space such that
\begin{eqnarray}
    \chi^2 ~=~ \left( \Delta \bm{M}_{\rm p} \right)^T \mathbf{C}_{\rm p} \left( \Delta \bm{M}_{\rm p} \right) \, .
    \label{chi_sq_Mp}
\end{eqnarray}
Applying the change of basis rule for matrices, we get that
\begin{eqnarray}
    \mathbf{C}_{\rm p} ~=~ \left( \mathbf{J}^{-1} \right)^T \mathbf{C}_{\rm obs} \mathbf{J}^{-1} \, ,
\end{eqnarray}
where $\mathbf{C}_{\rm obs}$ is the cost matrix in the plane of the $\bm{M}_{\rm lens}^{\rm Newton}$ observables. Assuming that strong lensing measurements at different apertures rely on strong lensing arc(s) at the corresponding projected radius, different arcs and thus different background sources must be used for each aperture. We therefore expect independent uncertainties on the two $M_{\rm lens}^{\rm Newton}$ values, making $\mathbf{C}_{\rm obs}$ diagonal. Each non-zero entry is then simply the corresponding inverse variance.
\begin{eqnarray}
    \mathbf{C}_{\rm obs} ~=~ \begin{bmatrix} \sigma^{-2} \left( M_{\rm lens, inner}^{\rm Newton} \right) & 0 \\
    0 & \sigma^{-2} \left( M_{\rm lens, outer}^{\rm Newton} \right) \end{bmatrix} \, ,
\end{eqnarray}
where $\sigma \left( X \right)$ is the uncertainty of $X$ for any quantity $X$, with inner/outer subscripts indicating the aperture.

Our approach does not necessarily lead to a symmetric cost matrix $\mathbf{C}_{\rm p}$. However, we only need it to work out $\chi^2$ based on the strong lensing observations. Equation~(\ref{chi_sq_Mp}) shows that we would get the same $\chi^2$ if we symmetrise $\mathbf{C}_{\rm p}$. We therefore set
\begin{eqnarray}
    \mathbf{C}_{\rm p} ~\to~ \frac{\mathbf{C}_{\rm p} + \mathbf{C}_{\rm p}^T}{2} \, .
\end{eqnarray}
We then diagonalise $\mathbf{C}_{\rm p}$ to find its eigenvalues and eigenvectors. Since $\mathbf{C}_{\rm p}$ is now symmetric, its two eigenvectors are orthogonal, greatly simplifying the interpretation. The eigenvectors define the two statistically independent directions in the $\bm{M}_{\rm p}$ plane. The corresponding eigenvalues give the inverse square of the distance one must go along each eigenvector direction to reach the point where $\chi^2 = 1$. This allows us to draw an error ellipse in the $\bm{M}_{\rm p}$ plane under the assumption that $\chi^2$ varies quadratically with position around the point corresponding to the observed $M_{\rm lens}^{\rm Newton}$ values.

We define the region centred on the northern BCG or central dominant (cD) galaxy of the main clump as the northern core, the region centred on the southern BCG as the southern core, and the region centred on the BCG of the subclump as the subclump core. \citetalias{Cha_2025} provides the GR strong lensing masses within 250~kpc of the three core regions. We obtain the corresponding GR strong lensing masses within 80, 100, and 150~kpc following the same method. Table~\ref{tab:GR_lensing_mass} lists the GR strong lensing masses that we obtain for the three core regions at different projected radii.

\setlength{\tabcolsep}{10pt}
\renewcommand{\arraystretch}{1.2}
\begin{table*}
    \centering
    \begin{tabular}{lcccc} 
    \hline
    GR strong lensing mass & $r_{\rm p}=$ 80 kpc & $r_{\rm p}=$ 100 kpc & $r_{\rm p}=$ 150 kpc & $r_{\rm p}=$ 250 kpc \\ \hline
    $M_{\rm GR,south}(<r_{\rm p})$ & $0.338 \pm 0.018$ & $0.475 \pm 0.035$ & $0.842 \pm 0.086$ & $1.670 \pm 0.140$ \\  
    $M_{\rm GR,north}(<r_{\rm p})$ & $0.338 \pm 0.012$ & $0.486 \pm 0.024$ & $0.898 \pm 0.070$ & $1.590 \pm 0.140$ \\
    $M_{\rm GR,sub}(<r_{\rm p})$ & $0.311 \pm 0.014$ & $0.409 \pm 0.025$ & $0.579 \pm 0.067$ & $0.870 \pm 0.130$ \\ \hline
    \end{tabular}
    \caption{The GR strong lensing mass ($M^{\rm Newton}_{\rm lens}$) within different sky-projected radii ($r_{\rm p}$) from the southern (south), northern (north), and subclump (sub) core. All masses are in units of $10^{14}\,M_{\odot}$. All GR strong lensing masses within the 250~kpc apertures are taken from \citetalias{Cha_2025}, while the lensing masses at all other radii are derived following their methodology. Hereafter, we collectively refer to all GR strong lensing masses in this table as the GR strong lensing masses from \citetalias{Cha_2025}.}
    \label{tab:GR_lensing_mass}
\end{table*}
\renewcommand{\arraystretch}{1}

These results allow us to constrain the MOND strong lensing masses. We find that the MOND predictions for the inner and outer $M_{\rm p}$ values are indeed correlated, but only to a small extent (see Appendix \ref{lensing_vs_bar}). We therefore mostly discuss our results as if there is no correlation between uncertainties in each predicted $M_{\rm p}$. For this, we project the eigenvectors onto the desired direction in the $\bm{M}_{\rm p}$ parameter space and combine uncertainties as if the eigenvector directions are statistically independent.

\section{The IGIMF theory and mass-to-light ratios}
\label{IGIMF}

The IGIMF theory suggests that stellar populations in galaxy clusters may be more massive than previously thought, contributing more mass than inferred using traditional assumptions about the IMF.
This is not, \emph{per se}, a consequence of the IGIMF theory, but rather of the high metallicities of massive early-type galaxies (ETGs), which can only be achieved if the galaxy-wide IMF has been top-heavy. The IGIMF theory merely provides the empirically-gauged theoretical formulation for the variation of the galaxy-wide IMF \citep{Yan_2021}.
The primary difference between the IGIMF theory and the canonical IMF lies in the top-heavy nature of the IGIMF during the early evolutionary stages of massive galaxies \citep{Kroupa_2001, Kroupa_2002_IMF}. The total $M_{\star}/L$ ratio of galaxies, including both living stars and stellar remnants, can be derived within the IGIMF framework using the stellar luminosity–mass relation, which allows us to estimate the total baryonic masses of the galaxies.

In this section, we briefly introduce how we calculate $M_\star/L$ based on the IGIMF theory, emphasizing its differences from the canonical IMF. We then present the $M_\star/L$ values of galaxies in different photometric bands under both the IGIMF and canonical IMF assumptions. For a detailed and comprehensive description as well as reviews of the observational evidence, we refer the reader to e.g., \citet{Jerabkova_2018, Yan_2021, Gjergo_2025_massive, Haslbauer_2024_IGIMF, Zonoozi_2025_SPS, Zonoozi_2025_BTFR, Kroupa_2026_IMF} and \citet{Zhang_2026PRD}.

A canonical IMF \citep[such as that proposed by][]{Kroupa_2001} characterizes the stellar mass distribution for individual star-forming events, typically within embedded clusters. It can be expressed as 
\begin{eqnarray}
    \xi_\star(m) ~=~ k_\star
    \begin{cases}
    2\, m^{-\alpha_1}, & 0.08\,M_\odot \le m < 0.5\,M_\odot, \\[6pt]
    m^{-\alpha_2},     & ~~0.5\,M_\odot \le m < 150\,M_\odot \, ,
    \end{cases}
    \label{MF1}
\end{eqnarray}
where $m$ is the mass of a star and $k_\star$ is the normalization constant. The number of stars formed in the mass interval $m$ to $m + \mathrm{d}m$ is given by $\mathrm{d}N_\star = \xi_\star(m)\,\mathrm{d}m$. The canonical stellar IMF has slopes $\alpha_1 = 1.3$ and $\alpha_2 = 2.3$ \citep[see][]{Kroupa_2026_IMF}. The factor of 2 ensures continuity if $\alpha_2 = \alpha_1 + 1$.

In the IGIMF framework, we adopt a stellar IMF that varies with the physical conditions of the molecular cloud clumps in which embedded stellar clusters form. For this framework, Eq.~(\ref{MF1}) can be rewritten as \citep{Kroupa_2001, Kroupa_2002_IMF, Yan_2020, Yan_2021, Haslbauer_2024_IGIMF,Gjergo_2025_massive}       
\begin{eqnarray}
    \xi_\star(m) ~=~ k_\star
    \begin{cases}
    2\, m^{-\alpha_1}, & 0.08\,M_\odot \le m < 0.5\,M_\odot, \\[6pt]
    m^{-\alpha_2},     & 0.5\,M_\odot \le m < 1.0\,M_\odot, \\[6pt]
    m^{-\alpha_3},     & 1.0\,M_\odot \le m < m_{\rm max} \, ,
    \end{cases}
    \label{MF2}
\end{eqnarray}
where $\alpha_1$ and $\alpha_2$ are functions of metallicity and $\alpha_3$ depends on both metallicity and the initial mass of the embedded cluster. $m_{\rm max} = m_{\rm max}(M_{\rm ecl})$ denotes the most massive star that forms in an embedded cluster with a total stellar mass of $M_{\rm ecl}$ \citep{Jerabkova_2018, Yan_2023, Zonoozi_2025_SPS}.

Most stars are born in embedded clusters. Regardless of whether they subsequently dissolve or remain gravitationally bound, embedded clusters contribute to the overall stellar population of the galaxy. Accordingly, the final gwIMF is defined as the sum of the individual stellar IMFs of all embedded clusters formed within a given time interval, $\delta t$. For the invariant canonical IMF form (Eq.~\ref{MF1}), this summation yields the same functional form. However, for the variable IMF case (under the IGIMF framework), we need to sum over the IMFs of all embedded clusters. With $\mathrm{d}N_{\rm ecl} = \xi_{\rm ecl}(M_{\rm ecl})\,\mathrm{d}M_{\rm ecl}$ representing the number of embedded clusters within the mass interval $[M_{\rm ecl},\, M_{\rm ecl} + \mathrm{d}M_{\rm ecl}]$, $\xi_{\rm ecl}(M_{\rm ecl})$ denotes the mass function of these embedded clusters, referred to as the Embedded Cluster Mass Function (ECMF). We assume it follows a single power-law distribution with a slope $\beta_\star$ that depends on the star formation rate (SFR) \citep{Recchi_2009, Weidner_2013, Yan_2017, Zhang_2018}:
\begin{equation}
    \xi_{\mathrm{ecl}}(M_{\mathrm{ecl}}) = 
    \begin{cases}
        0, & M_{\mathrm{ecl}} < M_{\mathrm{ecl,min}}, \\[2mm]
        k_{\mathrm{ecl}}\, M_{\mathrm{ecl}}^{-\beta_\star}, 
            & M_{\mathrm{ecl,min}} \le M_{\mathrm{ecl}} < M_{\mathrm{ecl,max}}(\mathrm{SFR}), \\[2mm]
        0, & M_{\mathrm{ecl}} \ge M_{\mathrm{ecl,max}}(\mathrm{SFR}) ,
    \end{cases}
    \label{MF333}
\end{equation}
where the slope
\begin{eqnarray}
    \beta_{*} ~=~ -0.106 \log_{10} \psi(t) + 2 \, ,
    \label{beta}
\end{eqnarray}
with $\psi(t)$ representing the SFR in units of $M_\odot/\rm{yr}$. The minimum stellar mass of embedded clusters is taken to be $M_{\rm ecl,min} = 5\,M_{\odot}$, following the smallest observed star-forming molecular cloud clumps \citep{Kroupa_2003_brown_dwarf, Kirk_2012, Joncour_2018}. The maximum mass of embedded clusters, $M_{\rm ecl,max}$, depends on the SFR. Both $M_{\rm ecl,max}$ and the normalization constant $k_{\rm ecl}$ are determined by simultaneously solving the following equations:
\begin{eqnarray}
    M_{\rm tot,\delta t} ~=~ \int_{M_{\rm ecl,min}}^{M_{\rm ecl,max}}\xi_{\rm ecl}(M)  M \mathrm{d}M ~&=&~ \psi(t)\delta t \, , \nonumber\\
    \int_{M_{\rm ecl,max}} ^{10^9 M_\odot} \xi_{\rm ecl}(M) \mathrm{d}M ~&=&~ 1 \, .
    \label{Mtot10}
\end{eqnarray}
Here, $M_{\rm tot,\delta t}$ denotes the total stellar mass formed in a galaxy during a single star-formation episode of duration $\delta t = 10$~Myr, corresponding to the typical lifetime of molecular clouds \citep{Kroupa_2013, Schulz_2015, Yan_2017, Kroupa_2026_IMF}. The second term in Eq.~(\ref{Mtot10}) indicates that a single cluster is formed within the interval between the maximum embedded cluster mass, $M_{\rm ecl,max}$, and the adopted upper cluster mass limit of $10^9\,M_{\odot}$. The results are insensitive to the exact choice of $M_{\rm ecl,max} > 10^9\,M_{\odot}$.

Galaxies are divided into two categories based on the presence or absence of ongoing star formation: spiral or late-type galaxies (LTGs) with active star formation, for which a constant star formation rate is assumed \citep{Kroupa_2020}; and elliptical or ETGs with quenched star formation, for which an initially very high star formation rate is adopted according to the downsizing relation \citep{Thomas_2005, McDermid_2015, Yan_2021}:
\begin{eqnarray}
    \frac{\tau_{_{\rm SF}}}{\rm Gyr} ~=~ 49 \left( \frac{M_{\star}}{M_{\odot}} \right)^{-0.14}\, ,
    \label{downsizing}
\end{eqnarray}
where $\tau_{_{\rm SF}}$ is the formation timescale of elliptical galaxies. This implies that more massive galaxies tend to form their stars earlier and over shorter timescales than less massive ones, essentially on a pre-galactic gas-cloud free-fall timescale \citep{Eappen_2022}.

We then construct the SPS model of galaxies using the \texttt{SPS-VarIMF} code \citep{Zonoozi_2025_SPS}. Following the relation described in \citet{Gallazzi_2005}, galaxies with masses ranging from $10^{5}\,M_{\odot}$ to $10^{13}\,M_{\odot}$ are assigned metallicities from $Z = 0.0002$ to $Z = 0.03$ \citep{Zonoozi_2025_SPS}. In the canonical IMF framework, metallicity evolution is not taken into account; as indicated in Eq.~(\ref{MF1}), the canonical IMF paradigm is not sensitive to metallicity. In the IGIMF framework, we consider two metallicity models: a constant-metallicity model similar to the canonical IMF case, and an idealized self-enriched metallicity-evolution model. In the latter case, we assume that all galaxies enrich their metallicities starting from $Z=0.0002$ and reach the metallicities assigned by \citet{Zonoozi_2025_SPS} at redshift $z=0$.

\begin{figure}
    \centering
    \includegraphics[width = \linewidth]{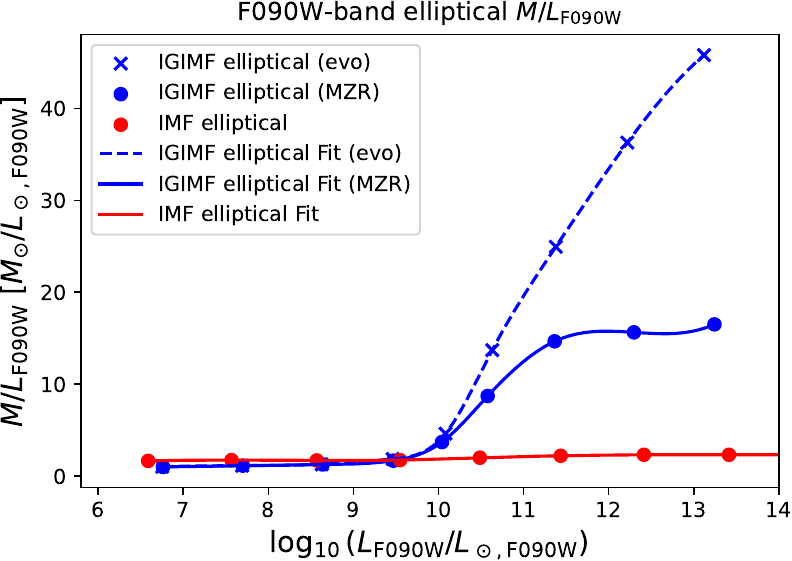}
    \includegraphics[width = \linewidth]{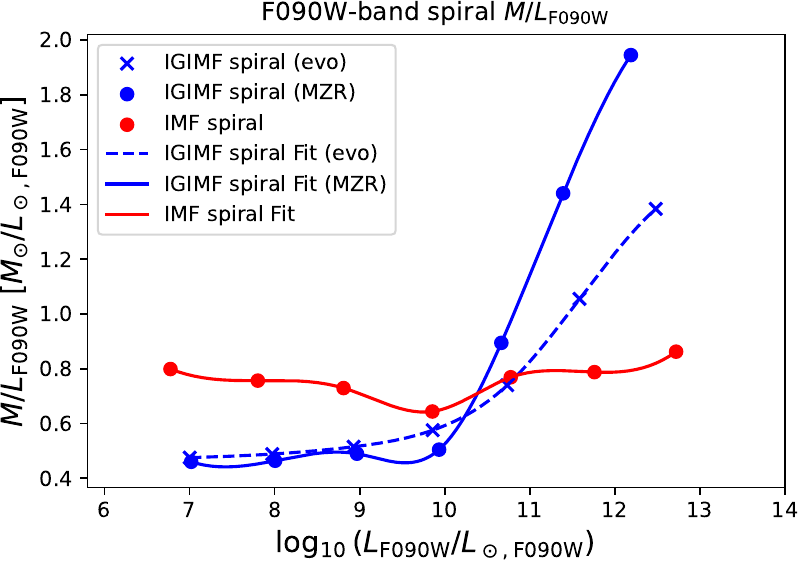}
    \caption{The relation between $M_\star/L$ and luminosity in the F090W band for elliptical galaxies (top panel) and spiral galaxies (bottom panel) for the canonical IMF (red dots) and the IGIMF (blue dots for the constant-metallicity model and blue crosses for the enriched-metallicity model). Note that $M_\star$ includes both stars and stellar remnants. The red solid line, blue solid line, and blue dashed line represent cubic-spline interpolations of the three corresponding results. Note the difference in the vertical scale between the two panels. Results for other photometric bands are shown in Appendix~\ref{M_L_ratios_appendix}.}
    \label{F090W}
\end{figure}

The SPS model simulates the evolutionary history of galaxies. From this, we extract the luminosities of galaxies with different morphologies (ETG or LTG) and masses at redshift $z = 0.296$ in various photometric bands, and thereby obtain the final $M_{\star}/L$. We calculate $M_{\star}/L$ in five JWST bands: F090W, F115W, F150W, F200W, and F277W. The corresponding results are shown in Figures~\ref{F090W}, \ref{F115W}, \ref{F150W}, \ref{F200W}, and \ref{F277W}, respectively. The $M_{\star}/L$ uncertainties are estimated from the error bounds of the cubic-spline interpolation \citep{Hall_1976}. From these five figures, it can be seen that the IGIMF predicts that massive elliptical galaxies have significantly higher $M_{\star}/L$ ratios than predicted by the canonical IMF. Furthermore, for the IGIMF case, the results from the enriched-metallicity model are larger than those from the constant-metallicity model. It should be noted that the luminosities of the BCGs in the Bullet Cluster are on the order of $10^{11}$--$10^{12} \, L_{\rm \odot, i}$, where $i$ denotes one of the aforementioned photometric bands. 
For spiral galaxies, the differences among the three models are small. The immediate reason is their low luminosities. In addition, spiral galaxies have star-formation histories that differ fundamentally from those of ellipticals and do not host a large population of stellar remnants.

The constant-metallicity model (following the observed mass-metallicity relation; MZR) keeps the metallicity fixed at its $z=0$ value, resulting in a higher mean metallicity and thus higher luminosity at the long wavelengths used in this work. On the other hand, the enriched-metallicity model (evolving; evo) begins with lower metallicity at early times, leading to a more top-heavy gwIMF and consequently a larger fraction of stellar remnants. These two effects combine to yield a higher $M_{\star}/L$ in the enriched-metallicity model compared with the constant-metallicity model.

To reproduce the observed super-solar metallicities of massive elliptical galaxies, they must have a top-heavy gwIMF in their early stages. This requirement has been discussed in a variety of studies \citep{Baugh_2005MNRAS, Smith_2020ARA&A, Van_Dokkum_2024ApJ}.
In the IGIMF theory, galaxies experience rapid and intense star formation at early times through the downsizing relation (Eq.~\ref{downsizing}), leading to a top-heavy gwIMF and consequently a very large fraction of stellar remnants. This explains why massive elliptical galaxies have higher $M_{\star}/L$ in the IGIMF framework than in the canonical IMF case. From a dynamical perspective, \cite{Dabringhausen_2008MNRAS} showed that the dynamical mass-to-light ratio increases with the dynamical mass of ETGs. This trend may indicate a larger contribution from stellar remnants in more massive ETGs. 

Nevertheless, we emphasize that testing the consistency between the IGIMF baryonic mass and the (MONDian) dynamical mass for massive ETGs lies beyond the scope of this work and remains to be explored.
For example, within the IGIMF framework, BCGs can attain very large baryonic masses. If this mass is concentrated toward the central regions, the system may be driven into a regime that is effectively Newtonian over most of its luminous extent, which could in principle lead to a declining or approximately Keplerian velocity-dispersion profile. This would appear to be in tension with the relatively flat radial velocity-dispersion profiles observed in BCGs \cite{Tian_2020ApJ,Tian_2021ApJ,Tian_2024A&A,Edwards_2024MNRAS}. It should be noted that the $M_{\star}/L$ inferred under the IGIMF framework corresponds to the global $M_{\star}/L$ of galaxies, rather than that measured within a limited aperture such as the effective radius. One possible way to mitigate the above tension is to consider that stellar remnants, which constitute a significant fraction of the baryonic mass, may develop a more extended spatial distribution than stars due to effects such as natal kicks. For instance, models from \citet{Sweeney_2022MNRAS} suggest that 30\% of stellar remnants in the Milky Way acquire sufficient kinetic energy from natal kicks to escape the Galaxy.

However, the relative mass fractions of black holes and neutron stars among stellar remnants depend on the shape of the gwIMFs in each galaxy, with the boundary mass between neutron stars and black holes depending also on the metallicity, spin of the precursor, and the presence of a companion, all of which result in some uncertainty. Since lower mass stars would form far more frequently and these leave behind neutron stars rather than black holes, it is quite plausible that neutron stars dominate the remnant population. In this case, the large number of associated supernova events would inject substantial amounts of metals and $\alpha$-elements (including O, Mg, Si, etc., which are related to the star formation timescale, SFT) into the intracluster medium (ICM). This may be in tension with the observed sub-solar metallicity of the ICM. Still, some studies suggest that the ICM in massive galaxy clusters may indeed require an additional early stellar population to account for the bulk of the metals \cite{Blackwell_2022,Blackwell_2026ApJ}.

On the other hand, if stellar-mass black holes constitute the dominant remnant component, the relatively low natal kick velocities expected from momentum conservation would lead to a reduced degree of spatial diffusion, thereby weakening the effectiveness of natal kicks in reconciling the dynamical mass discrepancy. Conversely, there are also arguments that black holes may receive kick velocities comparable to those of neutron stars \cite{Repetto_2012MNRAS,Repetto_2017MNRAS,Vanbeveren_2020A&A}.
In addition, some indirect studies, such as \cite{Peuten_2016MNRAS,Baumgardt_2017MNRAS,Pavlik_2018A&A}, suggest that about 50\% of black holes are retained in present-day-type globular clusters (GCs). GCs may have had a top-heavy initial mass function (IMF) \cite{Marks_2012}, implying that their initial masses could have been a factor of $10-100\times$ larger than currently inferred. In this case, larger natal kick velocities would be required for black holes to escape. Furthermore, the fraction of progenitors undergoing direct collapse or explosion \cite{Wirth_2024A&A}, as well as the metallicity, may also affect the distribution of kick velocities.
Another important issue concerns the lack of observational constraints on black holes. At present, most observational results for stellar-mass black holes are derived from binary systems, but black holes with large kick velocities are less likely to remain in binaries \citep{Nagarajan_2025PASP}. Future microlensing surveys may provide more robust constraints on the population of isolated black holes.

All these issues require a quantitative determination of the relative contributions of black holes and neutron stars to the stellar remnant population in order to be further clarified, as well as a quantitative dynamical study of the diffusion of stellar remnants.
Finally, the high $M_{\star}/L$ predicted by the IGIMF is a natural consequence of the high metallicity and high [$\alpha$/Fe] ratios exhibited by ETGs. If massive ETGs in the sample had low metallicities, deviating from the observed statistical trends, they would not exhibit significantly elevated $M_{\star}/L$. This issue ultimately requires more detailed observational constraints on the elemental abundances of galaxies.
A detailed discussion of the compatibility between the IGIMF and the dynamical mass inferred from MOND can be found in \cite{Zhang_2026PRD}.

\section{The mass of the core area of the Bullet Cluster}
\label{IGIMF-Bullet}

In this section, we estimate the total baryonic mass budget in the regions surrounding the northern and southern cD or BCG of the main clump, as well as the BCG of the subclump. We then compare the total baryonic mass budget with the MOND strong lensing mass. We use the JWST NIRCam imaging from GO-4598 (PI: Maru\v{s}a Brada\v{c}), which was employed in \citetalias{Cha_2025}, to construct the member-galaxy catalog of the Bullet Cluster. Since the mass reconstruction in \citetalias{Cha_2025} is based on a free-form lensing method \citep{Cha_2022, Cha_2023, Cha_2024} that does not rely on identifying cluster members, no redshift selection was imposed in the analysis. For the purpose of our baryonic mass estimation, we use the redshift information provided in the galaxy catalog and regard galaxies within $[z-\mathrm{d}z,\, z+\mathrm{d}z]$ as members, where $\mathrm{d}z = 0.06(1+z)$ and $z = 0.296$ is the Bullet Cluster redshift.

The morphological type of each member galaxy is determined based on three criteria:
\begin{enumerate}
    \item The short-to-long axis ratio > 0.5;
    \item The S\'{e}rsic index > 2; and
    \item $m_{\rm F606W} - m_{\rm F814W} > 1.2$, where $m_{\rm F606W}$ and $m_{\rm F814W}$ denote the observed apparent magnitudes in the Vega system in the F606W and F814W bands, respectively, without $K$-correction \citep{Zheng_2004}.
\end{enumerate}
A galaxy is classified as an early-type (elliptical) galaxy if at least two of these three criteria are satisfied; otherwise, it is classified as a late-type (spiral) galaxy.

Next, we estimate the luminosities of the member galaxies. The JWST catalog (processed with \texttt{Source Extractor}; \citealt{Bertin_1996}) provides the apparent magnitudes of the member galaxies measured within an aperture of 2.5 Kron radii, where the Kron radius is an adaptive elliptical aperture scale defined by the first moment of the light distribution \citep{Kron_1980}. To estimate the total luminosity of each galaxy, we assume that the light distribution follows a S\'{e}rsic profile and extrapolate the luminosity to infinity. This extrapolation has negligible effect for sources with a S\'{e}rsic index less than 1, but for typical elliptical galaxies with a S\'{e}rsic index greater than 2, it increases the total luminosity to approximately 120\% or more of the aperture luminosity \citep[this issue is discussed in detail in][]{Graham_2005}. We calculate the absolute magnitudes in the F090W, F115W, F150W, F200W, and F277W bands. We apply K-corrections following \citet{Manzoni_2025}, though the K-corrections are minor in these long-wavelength bands. The final stellar mass of each galaxy is taken as the average of the masses estimated from these five photometric bands.

We also obtain the ICL distribution in the F277W band. This dataset masks the light from all galaxies and foreground stars \citepalias[see also][]{Cha_2025}. We set the flux counts of all masked pixels to zero and sum the flux counts of the remaining pixels to obtain the ICL luminosity within different radial ranges centred on the three core regions. We then derive the ICL luminosity fraction $f_{\rm ICL} \equiv L_{\rm ICL} / (L_{\rm ICL} + L_{\rm galaxy})$ within each aperture based on the total galaxy luminosity in the F277W band. We estimate the ICL mass within each aperture using the mean mass-to-light ratio of all galaxies located within it.

\begin{figure}
    \centering
    \includegraphics[width = \linewidth]{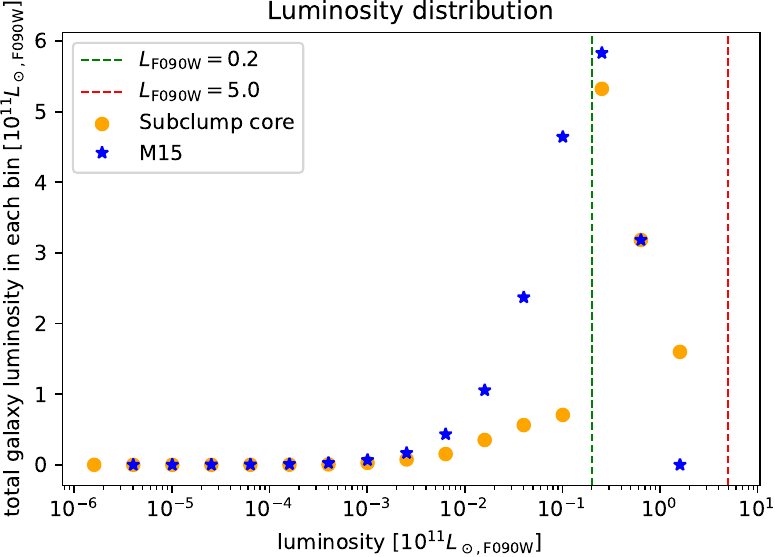}
    \caption{The orange dots show the F090W-band galaxy luminosity distribution within a projected radius of 250~kpc from the subclump core of the Bullet Cluster \citepalias{Cha_2025}, while the blue stars show the average luminosity function from a larger sample of galaxy clusters \citepalias{Moretti_2015}. The latter study is more complete due to the lower redshifts of its sample. The red and green dashed vertical lines indicate the luminosity range over which the Bullet Cluster observations are assumed to be complete. We therefore plot the orange points only after scaling them to match the galaxy luminosity function of \citetalias{Moretti_2015}, using the region between the dashed lines to determine the normalization for the Bullet Cluster and assuming that the overall distribution follows that of \citetalias{Moretti_2015}.}
    \label{Inc}
\end{figure}

Considering observational errors and the incompleteness of galaxy data due to the cosmological distance to the Bullet Cluster, we compare the catalog of its galaxies with the cluster-averaged Schechter luminosity function provided by \citet{Moretti_2015} (hereafter \citetalias{Moretti_2015}) to estimate the mass contribution of unobserved faint galaxies \citep{Meng_2024}. As an example, Fig.~\ref{Inc} shows a comparison between the average luminosity function from \citetalias{Moretti_2015} and the luminosity distribution of galaxies in the subclump core within a projected radius of 250~kpc, according to the JWST catalog. To account for the missing galaxies, we normalize the luminosity function based on the number of bright galaxies in the sample and extrapolate to fainter galaxies, assuming the same distribution of galaxy luminosities as in \citetalias{Moretti_2015}. The mass contributed by undetected galaxies inferred from this extrapolation is denoted as $M_{\rm faint}$, which we treat as an additional uncertainty that defines the upper limit of the total baryonic mass $M_{\rm bar}$ of the cluster.

\renewcommand{\arraystretch}{1.2}
\begin{table*}
    \centering
    \begin{tabular}{lcccc} 
    \hline
    Southern core & \multicolumn{4}{c}{Projected radius (kpc)} \\
    component & 80 kpc & 100 kpc & 150 kpc & 250 kpc \\ \hline
    $M_{\rm gas}$ & $0.079 \pm 0.026$ & $0.110 \pm 0.036$ & $0.270 \pm 0.089$ & $0.690 \pm 0.228$ \\  \hline
    $M_{\rm gal, IMF}$ & $0.012 \pm 0.000$ & $0.012 \pm 0.000$ & $0.020 \pm 0.000$ & $0.031 \pm 0.000$ \\ 
    $M_{\rm ICL, IMF}$ & $0.003 \pm 0.000$ & $0.004 \pm 0.000$ & $0.007 \pm 0.000$ & $0.010 \pm 0.000$ \\ 
    $M_{\rm faint, IMF}$ & $0.002$ & $0.003$ & $0.004$ & $0.013$ \\ 
    $M_{\rm bar, IMF}$ & $0.094 \pm 0.026$ & $0.127 \pm 0.036$ & $0.297 \pm 0.089$ & $0.731 \pm 0.228$ \\ \hline
    $M_{\rm gal, IG}$ & $[0.062, 0.172]$ & $[0.063, 0.174]$ & $[0.089, 0.236]$ & $[0.132, 0.344]$  \\ 
    $\Delta M_{\rm gal, IG}$ & $[0.002, 0.008]$ & $[0.002, 0.008]$ & $[0.002, 0.009]$ & $[0.003, 0.010]$  \\ 
    $M_{\rm ICL, IG}$ & $[0.018, 0.059]$ & $[0.023, 0.073]$ & $[0.032, 0.100]$ & $[0.046, 0.141]$  \\ 
    $\Delta M_{\rm ICL, IG}$ & $[0.002, 0.006]$ & $[0.002, 0.007]$ & $[0.002, 0.008]$ & $[0.002, 0.008]$  \\ 
    $M_{\rm faint, IG}$ & $[0.004, 0.009]$ & $[0.006, 0.014]$ & $[0.007, 0.017]$ & $[0.030, 0.061]$  \\ 
    $M_{\rm bar, IG}$ & $[0.159, 0.310]$ & $[0.196, 0.357]$ & $[0.391, 0.606]$ & $[0.868, 1.175]$  \\ 
    $\Delta M_{\rm bar, IG}$ & $\left[^{0.026}_{0.026}, \ ^{0.029}_{0.028}\right]$ & $\left[^{0.037}_{0.036}, \ ^{0.040}_{0.037}\right]$ & $\left[^{0.089}_{0.089}, \ ^{0.091}_{0.090}\right]$ & $\left[^{0.230}_{0.228}, \ ^{0.236}_{0.228}\right]$  \\  \hline 
    $M_{\rm lens, MOND}$ & $0.210 \pm 0.017$ & $0.288 \pm 0.033$ & $0.479 \pm 0.086$ & $0.862 \pm 0.089$ \\  \hline
    \end{tabular}
    \caption{The components of the 2D projected mass in units of $10^{14}\,M_{\odot}$ within different radii of the southern core. $M_{\rm gal,IMF}$, $M_{\rm ICL,IMF}$, and $M_{\rm faint,IMF}$ denote the stellar mass of galaxies, the ICL mass, and the mass of faint galaxies, respectively, all estimated using the mass-to-light ratio under the canonical IMF framework. The total baryonic mass in this case is defined as $M_{\rm bar,IMF} = M_{\rm gal,IMF} + M_{\rm ICL,IMF} + M_{\rm gas}$, where $M_{\rm faint,IMF}$ is considered only as an additional uncertainty to $M_{\rm bar,IMF}$. The subscript ``IG'' indicates the quantities derived under the IGIMF framework, but the definitions remain the same as above. $1\sigma$ uncertainties are shown here after $\pm$ symbols. The bottom of the interval corresponds to the result from the constant-metallicity (MZR) model, while the top corresponds to the enriched-metallicity (evo) model. $M_{\rm lens,MOND}$ denotes the MOND strong lensing mass derived using the method described in Section~\ref{Relativistic}, based on the GR strong lensing masses listed in Table~\ref{tab:GR_lensing_mass}. We adopt the 250~kpc mass from the 150 vs 250~kpc comparison as the final 250~kpc MOND strong lensing mass.}
    \label{Mass_budget}
\end{table*}
\renewcommand{\arraystretch}{1}

Finally, for the X-ray gas mass, we adopt the 2D distribution model from \citet{Paraficz_2016} (hereafter \citetalias{Paraficz_2016}), normalizing it with their provided X-ray gas mass within 150~kpc of the southern core in order to estimate the X-ray gas mass in both the main and subclumps. The results for the southern, northern, and subclump cores are summarized in Tables~\ref{Mass_budget}, \ref{Mass_budget_north}, and \ref{Mass_budget_sub}, respectively.

\begin{figure}
    \centering
    \includegraphics[width = \linewidth]{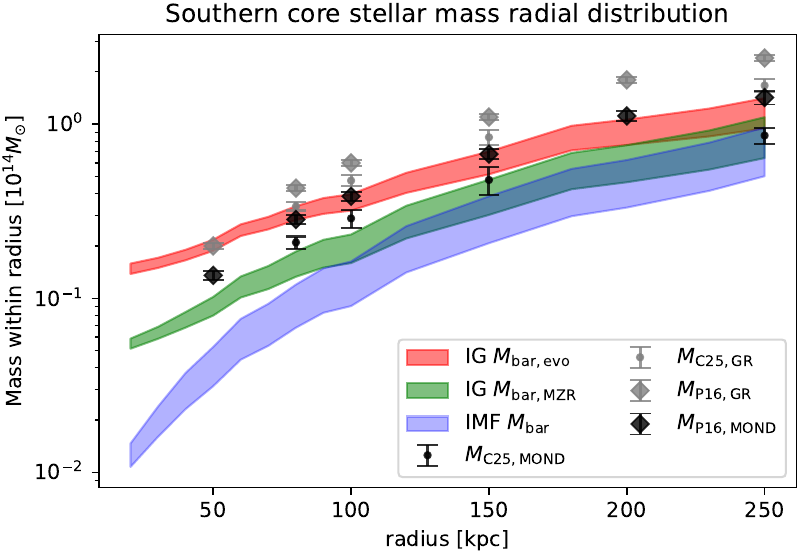}
    \includegraphics[width = \linewidth]{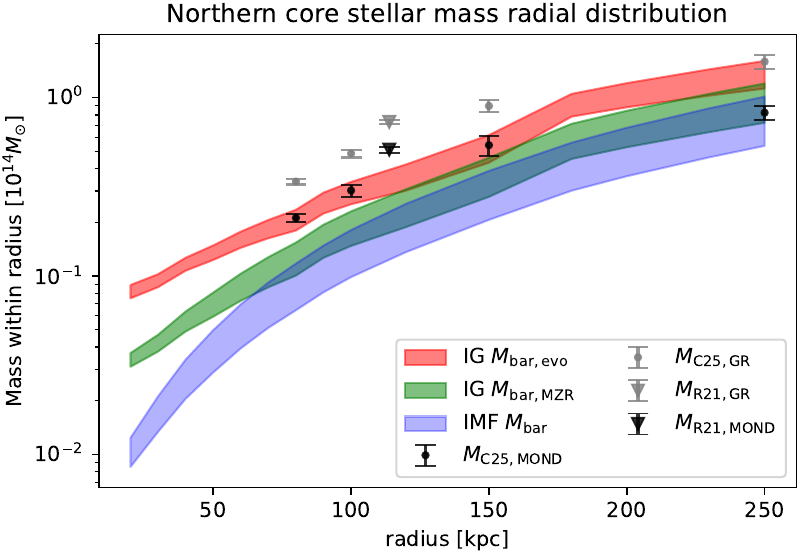}
    \includegraphics[width = \linewidth]{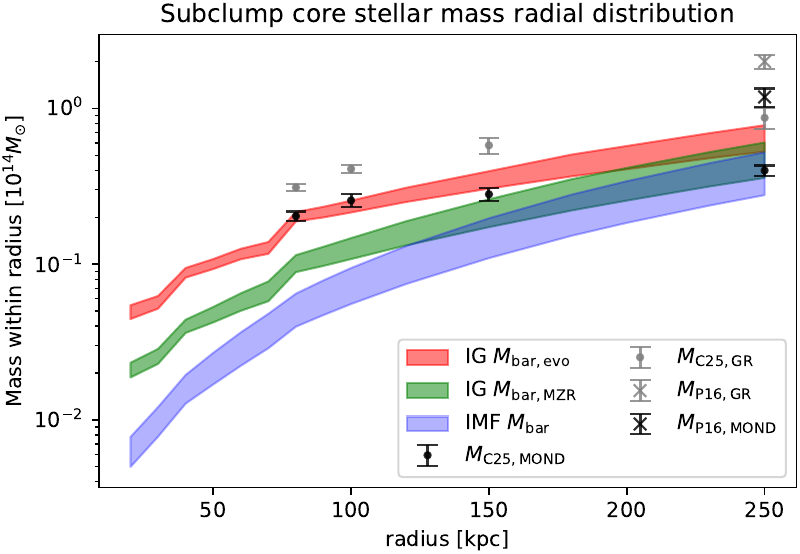}
    \caption{2D projected total baryonic and strong lensing mass distributions. From top to bottom, the panels are centred on the southern BCG, the northern cD/BCG, and the subclump BCG, respectively. The total baryonic mass includes the stellar mass of galaxies, the ICL mass, and the gas mass. The green and red shaded regions represent the $1\sigma$ uncertainties of the IGIMF baryonic masses under the enriched-(evo) and constant-metallicity (MZR) models, respectively. The blue shaded region corresponds to the $1\sigma$ uncertainty of the baryonic mass under the canonical IMF. The grey dots with $1\sigma$ error bars indicate the GR strong lensing masses obtained from \citetalias{Cha_2025}, while the grey triangle marks the GR strong lensing mass within about 113~kpc of the northern core reported by \citetalias{Richard_2021}. The grey diamonds and cross represent the GR strong lensing masses of the southern core and the subclump core, respectively, obtained from \citetalias{Paraficz_2016}. The markers of the corresponding shape in black, with error bars, show the MOND strong lensing masses calculated from these GR results following Section~\ref{Relativistic}.}
    \label{baryon_mass}
\end{figure}

In Fig.~\ref{baryon_mass}, we show the 2D projected distributions of the total baryonic mass within 250~kpc for the three core regions, along with the GR strong lensing masses reported in \citetalias{Paraficz_2016}, \citet{Richard_2021} (hereafter \citetalias{Richard_2021}), and \citetalias{Cha_2025}. We also show the corresponding MOND strong lensing masses. From Fig.~\ref{baryon_mass}, it is evident that the 2D projected baryonic mass distributions under the IGIMF framework agree well with the MOND strong lensing masses derived from the GR strong lensing masses reported in \citetalias{Cha_2025}. The lensing mass estimates based on earlier studies (\citetalias{Paraficz_2016,Richard_2021}) were all higher to varying degrees.

At larger radii, the MOND strong lensing masses tend to be closer to the predictions of the constant-metallicity model. Expressed as a fraction of the corresponding MOND strong lensing masses, the total baryonic masses under the IGIMF framework within 250~kpc are $\left[101^{+27}_{-26}\%, 136^{+27}_{-26}\%\right]$, $\left[117^{+29}_{-29}\%, 165^{+30}_{-29}\%\right]$, and $\left[120^{+31}_{-31}\%, 163^{+32}_{-31}\%\right]$ for the southern, northern, and subclump cores, respectively. The bottom (top) of the interval corresponds to the IGIMF mass obtained from the constant (evolving) metallicity model. As explained in Section~\ref{Discussion}, we consider these to be the lower and upper limits, respectively, of the IGIMF estimated masses. If instead we express each IGIMF mass as a fraction of the GR strong lensing mass from \citetalias{Cha_2025}, the IGIMF-based baryonic masses are $\left[52^{+14}_{-14}\%, 70^{+14}_{-14}\%\right]$, $\left[60^{+15}_{-15}\%, 86^{+15}_{-15}\%\right]$, and $\left[55^{+14}_{-14}\%, 75^{+15}_{-14}\%\right]$ for each of the cores. Thus, the required dark matter fraction is reduced. However GR is still unable to explain the observed strong lensing using the available baryons, in contrast to the above MOND + IGIMF results.

\section{Discussion}
\label{Discussion}

Our most important result is that the MOND strong lensing masses lie between the masses predicted by the IGIMF models with constant and enriched metallicities. As described in Section~\ref{IGIMF}, within the IGIMF framework, the constant-metallicity model should be regarded as a lower limit, since real galaxies are expected to experience metallicity enrichment during their evolution. 
In contrast, the idealized enriched-metallicity model should be regarded as an upper limit for the following reasons:
\begin{enumerate}
    \item The SFTs of elliptical galaxies used in this work follow the empirical downsizing relation of Eq.~(\ref{downsizing}) \citep{Yan_2021}, which is based on the results of \citet{Thomas_2005} and \citet{McDermid_2015}. In reality, however, galaxies may experience longer and more complex SFTs due to environmental processes, internal feedback, or continued gas accretion. Longer SFTs tend to produce lower average stellar metallicities and younger stellar populations, both of which act to reduce $M_{\star}/L$. Thus, the enriched-metallicity model, which assumes short star formation, naturally yields an upper bound on $M_{\star}/L$.
    \item The model further assumes that galaxies evolve as isolated systems forming through a pure downsizing sequence. Although this is supported as a dominant formation channel for massive ellipticals, real galaxies may also undergo mergers, tidal interactions, or extended gas infall. These processes generally prolong star formation and modify the pathway of metallicity enrichment, leading to effectively lower metallicities and lower $M_{\star}/L$ than predicted by an isolated downsizing model. Moreover, our implementation adopts a closed-box chemical evolution starting from low initial metallicity, which maximizes the efficiency of metal enrichment; in such a scenario, the IGIMF becomes bottom-heavy at early times, producing a higher $M_{\star}/L$. Real galaxies, however, may lose metals through galactic winds or acquire low-metallicity gas through inflows, both of which dilute the metallicity and suppress the resulting $M_{\star}/L$.
\end{enumerate}

A more accurate IGIMF mass-to-light ratio should be established individually for each member galaxy based on its own metallicity and [$\alpha$/Fe] ratio. However, no such observational data are currently available, so a more precise estimation of galaxy masses must await future observational progress in this field. We also find that the MOND strong lensing masses at small radii near the BCGs are closer to the baryonic masses predicted by the enriched-metallicity model. A possible explanation is that at small radii, a few massive elliptical galaxies, including the BCGs, dominate the total baryonic mass, and their evolutionary histories are more consistent with the enriched-metallicity scenario, leading to a higher fraction of stellar remnants compared with the constant-metallicity model. In contrast, at 250~kpc, the MOND strong lensing masses are closer to the baryonic masses derived from the constant-metallicity model. This may be because at larger radii, the region contains a larger number of intermediate- and low-mass galaxies, whose mass estimates are better described by a constant-metallicity model. The X-ray gas mass also becomes more significant at larger radii, making uncertainties in it relatively more important.

We note that our results are susceptible to observational limitations and inaccuracies. For instance, \citet{Jan_2025} pointed out that the fact that the hot gas is not in dynamical equilibrium can lead to overestimations in the X-ray mass. Moreover, the ICL still remains an active topic of research, with both its definition and observational techniques still under development \citep[see, e.g.,][]{Contini_2021}. In our estimation, the ICL luminosity fraction accounts for about 20--30\% of the total luminosity, similar to the result of \citet{Teja_2018}. This fraction is obtained by masking all member galaxies and foreground objects, a method that effectively neglects the ICL component overlapping with the profiles of member galaxies, particularly the BCGs. Modelling approaches that reconstruct the ICL distribution may yield different results \citep{Montes_2022}. On the other hand, we assume that the mass-to-light ratio of the ICL is equal to the average value of the galaxies within the same region, which may lead to a potential underestimation. Recent studies have suggested that the ICL was already enriched at $z>1$ \citep{Joo_2023, Werner_2023}, and its mass-to-light ratio may have been underestimated \citep{Tang_2018}. \citet{Joo_2023} proposed a scenario in which the majority of the ICL is assembled simultaneously with the growth of the BCG, or via the accretion of pre-processed intracluster stars, rather than being primarily generated through gradual stripping processes.

Using the GR strong lensing masses provided by \citetalias{Cha_2025}, we find that the resulting MOND strong lensing masses fall within the baryonic mass budget inferred under the IGIMF framework. In contrast, earlier studies \citepalias{Paraficz_2016, Richard_2021} reported higher GR strong lensing masses, which in turn led to higher MOND strong lensing masses. In \citetalias{Cha_2025}, the authors noted this discrepancy and emphasized the need for future spectroscopic redshift data to better constrain and correct the GR strong lensing masses. A GR lensing analysis of the Bullet Cluster will be presented in \cite{Cho_2025}, based on a joint JWST and wide-field DECam dataset.

Finally, as discussed in Section~\ref{IGIMF}, our work should at present be regarded as an estimate of the baryonic mass budget in the core regions of the Bullet Cluster, reflecting the potentially available baryonic mass in these regions. The consistency between the IGIMF framework and the (MONDian) dynamical masses of massive ETGs remains to be studied.

\section{Conclusions} 
\label{Conclusions}

The Bullet Cluster has long been a cornerstone in the study of dark matter, providing supposedly compelling evidence for an unseen mass component based on the observed offset between baryonic gas and the gravitational mass inferred from weak lensing. However, some assumptions in the analysis have been questioned, including the hydrostatic equilibrium of the hot gas. Moreover, the extreme properties of the Bullet Cluster and similar systems like El Gordo \citep*{Asencio_2021, Asencio_2023} pose a significant challenge for $\Lambda$CDM and motivate consideration of alternative frameworks such as MOND. In this work, we re-examine the compatibility of the MOND paradigm with the core regions of the Bullet Cluster under the IGIMF framework.

Our analysis of the Bullet Cluster focused on its gravitational field inferred through strong lensing. In MOND, we expect a GR-like relation between the gravitational field and the light deflection angle (Equation~\ref{Light_deflection_angle}). The mass required to explain the observed strong lensing is lower in MOND than in GR because of the enhanced gravity. The required lensing mass in GR and MOND is shown in Table~\ref{tab:GR_lensing_mass} and in the bottom part of Tables~\ref{Mass_budget}, \ref{Mass_budget_north}, and \ref{Mass_budget_sub} within four projected radii according to the observational studies in \citetalias{Cha_2025}.

To assess whether the MOND lensing mass is compatible with the baryonic mass budget, we estimate the latter using the IGIMF theory, which is required to explain the high metallicities in elliptical galaxies. It has also proven successful in explaining the dynamics of the ultra-diffuse galaxy Dragonfly 44 \citep{Haghi_2019_DF44} and the baryonic Tully-Fisher relation \citep[BTFR;][]{Zonoozi_2025_BTFR}. Furthermore, it naturally accounts for the top-light and top-heavy galaxy-wide IMFs deduced for dwarf and massive disc galaxies, respectively \citep{Lee_2009, Gunawardhana_2011, Kroupa_2026_IMF}. By incorporating the IGIMF’s prediction of a top-heavy gwIMF in the early stages of massive elliptical galaxies and including the contribution of the ICL, we estimated the total stellar mass, which we then combined with the estimated gas mass to obtain the total baryonic mass in the three core regions of the Bullet Cluster.

Our results show that, at all examined radii, the MOND strong lensing masses derived from the GR strong lensing masses of \citetalias{Cha_2025} lie between the lower and upper limits of the IGIMF-based baryonic masses. Considering that the actual evolutionary histories of galaxies must lie between the constant- and idealized enriched-metallicity models, this implies that, based on current observations, MOND appears to be broadly consistent with the baryonic mass budget in the core regions of the Bullet Cluster within the IGIMF framework. Future quantitative tests of the consistency with dynamical mass, together with observations of the chemical abundances of individual galaxies, will allow for tighter constraints on the stellar masses predicted by the IGIMF. In addition, microlensing observations hold the potential to more accurately constrain the fraction and spatial distribution of stellar remnants, providing an independent test of their predicted high abundance. In contrast, it is already clear that when compared with the GR strong lensing masses, a significant component of missing mass still remains: GR cannot provide enough gravity from the available baryons. This discrepancy is worsened if we assume an invariant gwIMF, which in addition is unable to account for the high metallicities of elliptical galaxies. Meanwhile, dark matter models also face the need to revise their parameters to reduce the required dark matter mass in order to align with the total baryonic mass predicted by the IGIMF theory.

In conclusion, the notion that the Bullet Cluster provides definitive evidence for dark matter needs to be questioned. Alternative frameworks, particularly MOND and the IGIMF theory, offer explanations that deserve further exploration. It is important to emphasize that the IGIMF framework has been developed since \citet{Kroupa_2003_IGIMF} without any relation to the missing gravity problem. It is based on observationally gauging the variation of the stellar IMF on the molecular clump scale based on detailed star count, dynamical, and astrophysical investigations of globular clusters, ultra-compact dwarf galaxies, and Galactic field stellar populations, leading to a formulation of how the stellar IMF depends on metallicity and density \citep{Dabringhausen_2009, Dabringhausen_2012, Marks_2012, Jerabkova_2018,Gjergo_2025_massive}. The IGIMF theory has thus not been constructed to remove the missing gravity problem in MOND, but does so for the Bullet Cluster. This is due to the high prevalence of elliptical galaxies and their rapid star formation timescales, which leads to a high proportion of massive stars that leave behind dark stellar remnants at late times.

\section*{Acknowledgements}

This work is based on observations taken by the NASA/ESA Hubble Space Telescope and NASA/ESA/CSA James Webb Space Telescope, downloaded from the Mikulski Archive for Space Telescope (MAST) at the Space Telescope Science Institute (STScI). DZ thanks the China Scholarship Council (CSC) for financial support. HH is grateful to the staff at the Helmholtz-Institut f{\"u}r Strahlen- und Kernphysik (HISKP) and Argelander Institut f{\"u}r Astronomie (AIfA) for their hospitality and acknowledges financial support from the Stellar Populations and Dynamics (SPODYR) group at the University of Bonn. EA acknowledges support through a teaching assistantship by the HISKP. IB is supported by Royal Society University Research Fellowship grant 211046. AHZ acknowledges support from the Alexander von Humboldt Foundation. PK acknowledges support through grant 26-21774S from the Czech Grant Agency and also through the DAAD Eastern-European Exchange Scheme between Bonn and Prague. AL acknowledges funding by the Deutsche Forschungsgemeinschaft (DFG) Research Training Group 2696. IB thanks Benoit Famaey for helpful comments regarding lensing in MOND. The authors also thank Danuta Paraficz for helpful assistance.




\bibliography{ref.bib} 

\begin{appendix}

\section{Mass estimates}
\label{Mass_estimates_appendix}

In this section, we present the baryonic mass estimates for the canonical IMF and the IGIMF as well as the MOND strong lensing masses. These are shown for the Bullet Cluster's northern core (Table~\ref{Mass_budget_north}) and subclump core (Table~\ref{Mass_budget_sub}).

\renewcommand{\arraystretch}{1.2}
\begin{table*}
    \centering
    \begin{tabular}{lcccc} 
    \hline
    Northern core & \multicolumn{4}{c}{Projected radius (kpc)} \\
    component & 80 & 100 & 150 & 250 \\ \hline
    $M_{\rm gas}$ & $0.080 \pm 0.027$ & $0.125 \pm 0.041$ & $0.276 \pm 0.091$ & $0.721 \pm 0.238$ \\  \hline
    $M_{\rm gal, IMF}$ & $0.008 \pm 0.000$ & $0.012 \pm 0.000$ & $0.016 \pm 0.000$ & $0.043 \pm 0.000$ \\ 
    $M_{\rm ICL, IMF}$ & $0.003 \pm 0.000$ & $0.003 \pm 0.000$ & $0.006 \pm 0.000$ & $0.010 \pm 0.000$ \\ 
    $M_{\rm faint, IMF}$ & $0.001$ & $0.002$ & $0.004$ & $0.011$ \\ 
    $M_{\rm bar, IMF}$ & $0.091^{0.027}_{0.027}$ & $0.140^{0.041}_{0.041}$ & $0.297^{0.091}_{0.091}$ & $0.774^{0.238}_{0.238}$ \\ \hline
    $M_{\rm gal, IG}$ & $[0.034, 0.089]$ & $[0.049, 0.125]$ & $[0.068, 0.172]$ & $[0.192, 0.499]$  \\ 
    $\Delta M_{\rm gal, IG}$ & $[0.001, 0.005]$ & $[0.001, 0.005]$ & $[0.002, 0.006]$ & $[0.003, 0.012]$  \\ 
    $M_{\rm ICL, IG}$ & $[0.012, 0.038]$ & $[0.015, 0.045]$ & $[0.025, 0.076]$ & $[0.047, 0.142]$  \\ 
    $\Delta M_{\rm ICL, IG}$ & $[0.001, 0.004]$ & $[0.001, 0.004]$ & $[0.001, 0.005]$ & $[0.002, 0.007]$  \\ 
    $M_{\rm faint, IG}$ & $[0.003, 0.008]$ & $[0.004, 0.009]$ & $[0.008, 0.018]$ & $[0.022, 0.045]$  \\ 
    $M_{\rm bar, IG}$ & $[0.127, 0.207]$ & $[0.189, 0.294]$ & $[0.369, 0.523]$ & $[0.960, 1.362]$  \\ 
    $\Delta M_{\rm bar, IG}$ & $\left[^{0.027}_{0.027}, \ ^{0.028}_{0.027}\right]$ & $\left[^{0.042}_{0.041}, \ ^{0.043}_{0.042}\right]$ & $\left[^{0.091}_{0.091}, \ ^{0.093}_{0.091}\right]$ & $\left[^{0.239}_{0.238}, \ ^{0.243}_{0.238}\right]$  \\  \hline
    $M_{\rm lens, MOND}$ & $0.212 \pm 0.012$ & $0.301 \pm 0.023$ & $0.541 \pm 0.070$ & $0.823 \pm 0.070$ \\  \hline
    \end{tabular}
    \caption{Same as Table~\ref{Mass_budget}, but for the northern core.}
    \label{Mass_budget_north}
\end{table*}
\renewcommand{\arraystretch}{1}

\renewcommand{\arraystretch}{1.2}
\begin{table*}
    \centering
    \begin{tabular}{lcccc} 
    \hline
    Subclump core & \multicolumn{4}{c}{Projected radius (kpc)} \\
    component & 80 & 100 & 150 & 250 \\ \hline
    $M_{\rm gas}$ & $0.038 \pm 0.012$ & $0.059 \pm 0.019$ & $0.133 \pm 0.044$ & $0.370 \pm 0.122$ \\  \hline
    $M_{\rm gal, IMF}$ & $0.012 \pm 0.000$ & $0.013 \pm 0.000$ & $0.016 \pm 0.000$ & $0.023 \pm 0.000$ \\ 
    $M_{\rm ICL, IMF}$ & $0.002 \pm 0.000$ & $0.003 \pm 0.000$ & $0.004 \pm 0.000$ & $0.006 \pm 0.000$ \\ 
    $M_{\rm faint, IMF}$ & $0.002$ & $0.002$ & $0.004$ & $0.013$ \\ 
    $M_{\rm bar, IMF}$ & $0.052^{0.013}_{0.012}$ & $0.075^{0.020}_{0.019}$ & $0.153^{0.044}_{0.044}$ & $0.399^{0.123}_{0.122}$ \\ \hline
    $M_{\rm gal, IG}$ & $[0.053, 0.134]$ & $[0.056, 0.139]$ & $[0.065, 0.163]$ & $[0.086, 0.212]$  \\ 
    $\Delta M_{\rm gal, IG}$ & $[0.001, 0.005]$ & $[0.001, 0.005]$ & $[0.001, 0.005]$ & $[0.002, 0.005]$  \\ 
    $M_{\rm ICL, IG}$ & $[0.010, 0.030]$ & $[0.013, 0.037]$ & $[0.019, 0.054]$ & $[0.024, 0.068]$  \\ 
    $\Delta M_{\rm ICL, IG}$ & $[0.001, 0.002]$ & $[0.001, 0.003]$ & $[0.001, 0.003]$ & $[0.001, 0.003]$  \\ 
    $M_{\rm faint, IG}$ & $[0.003, 0.006]$ & $[0.003, 0.006]$ & $[0.006, 0.011]$ & $[0.021, 0.042]$  \\ 
    $M_{\rm bar, IG}$ & $[0.101, 0.201]$ & $[0.127, 0.235]$ & $[0.217, 0.350]$ & $[0.480, 0.650]$  \\ 
    $\Delta M_{\rm bar, IG}$ & $\left[^{0.013}_{0.013}, \ ^{0.015}_{0.013}\right]$ & $\left[^{0.020}_{0.019}, \ ^{0.021}_{0.020}\right]$ & $\left[^{0.044}_{0.044}, \ ^{0.045}_{0.044}\right]$ & $\left[^{0.124}_{0.122}, \ ^{0.129}_{0.122}\right]$  \\  \hline
    $M_{\rm lens, MOND}$ & $0.203 \pm 0.014$ & $0.256 \pm 0.025$ & $0.281 \pm 0.027$ & $0.400 \pm 0.030$ \\  \hline
    \end{tabular}
    \caption{Same as Table~\ref{Mass_budget}, but for the subclump core.}
    \label{Mass_budget_sub}
\end{table*}
\renewcommand{\arraystretch}{1}

\section{Mass-to-light ratios}
\label{M_L_ratios_appendix}

\begin{figure}
    \centering
    \includegraphics[width = \linewidth]{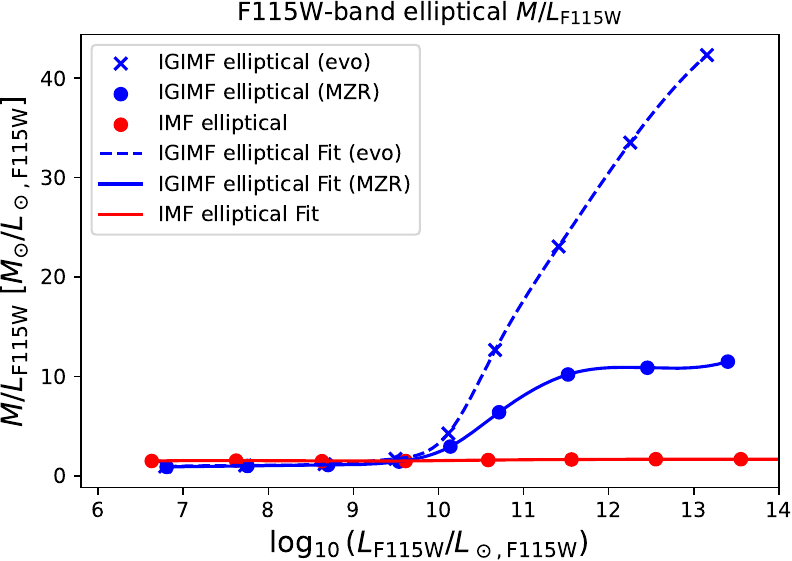}
    \includegraphics[width = \linewidth]{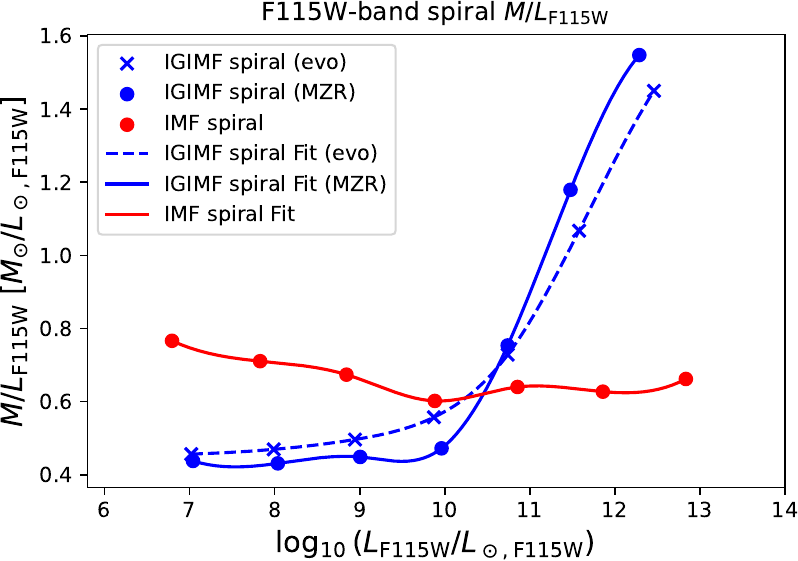}
    \caption{Same as Fig.~\ref{F090W}, but for the F115W band.}
    \label{F115W}
\end{figure}

\begin{figure}
    \centering
    \includegraphics[width = \linewidth]{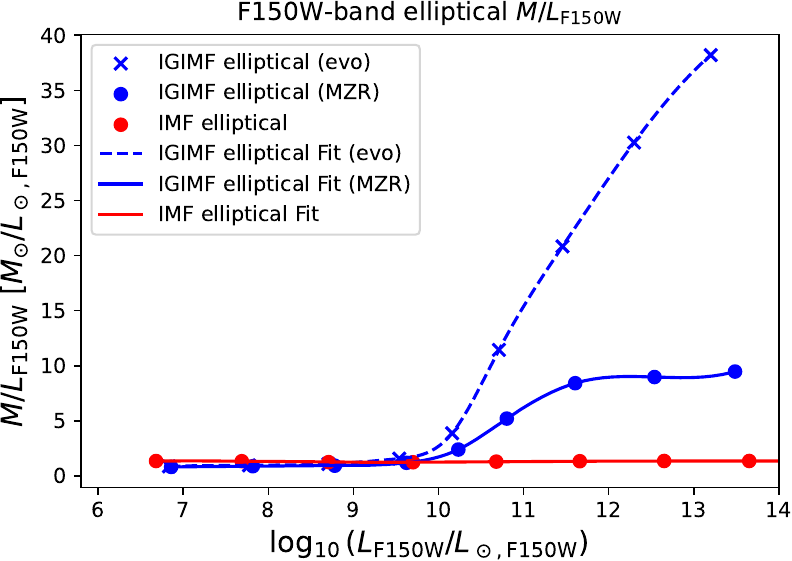}
    \includegraphics[width = \linewidth]{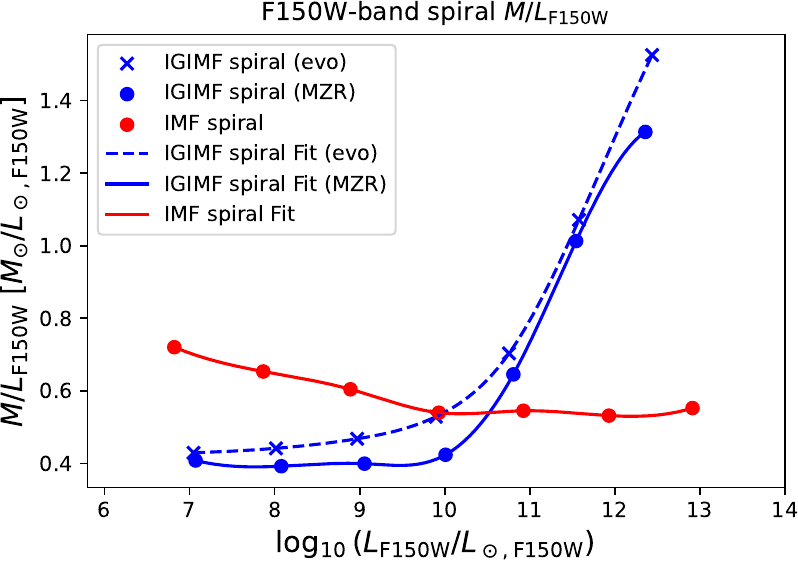}
    \caption{Same as Fig.~\ref{F090W}, but for the F150W band.}
    \label{F150W}
\end{figure}

\begin{figure}
    \centering
    \includegraphics[width = \linewidth]{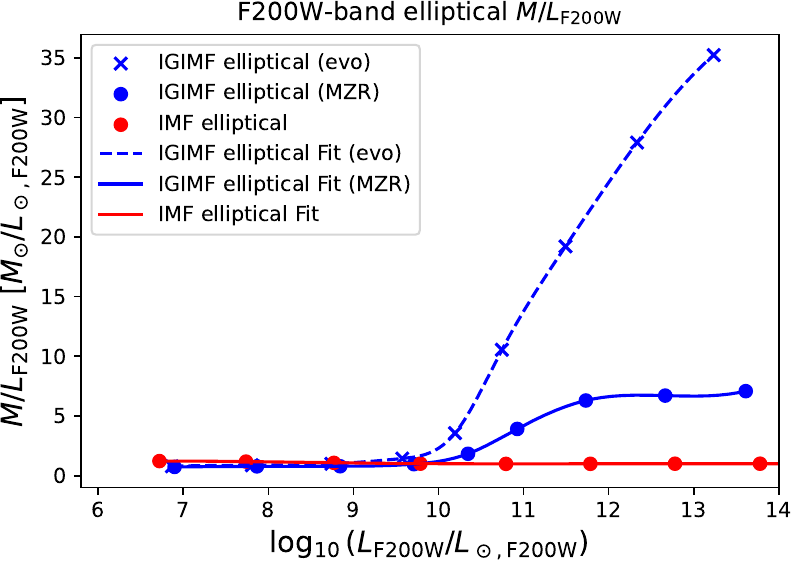}
    \includegraphics[width = \linewidth]{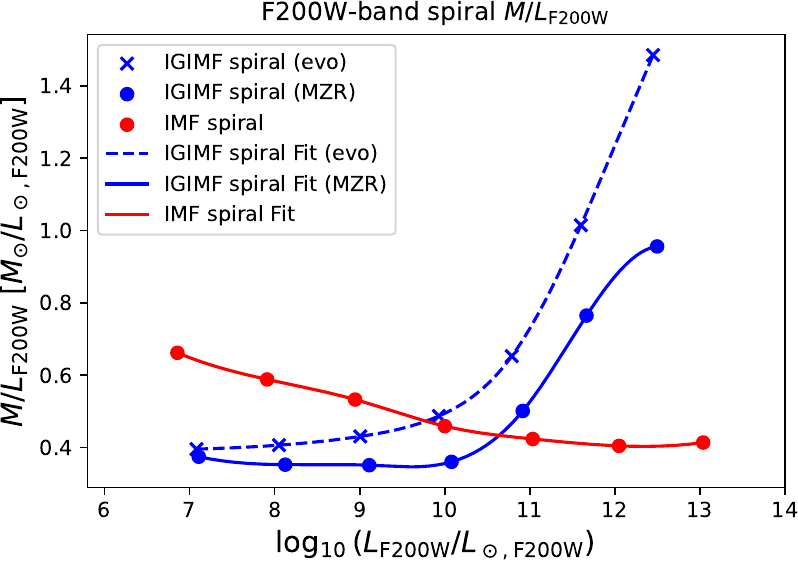}
    \caption{Same as Fig.~\ref{F090W}, but for the F200W band.}
    \label{F200W}
\end{figure}

\begin{figure}
    \centering
    \includegraphics[width = \linewidth]{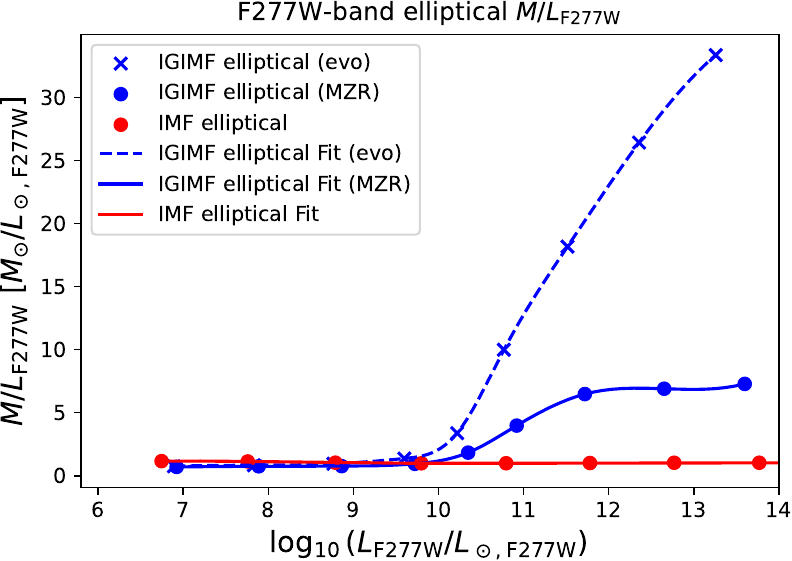}
    \includegraphics[width = \linewidth]{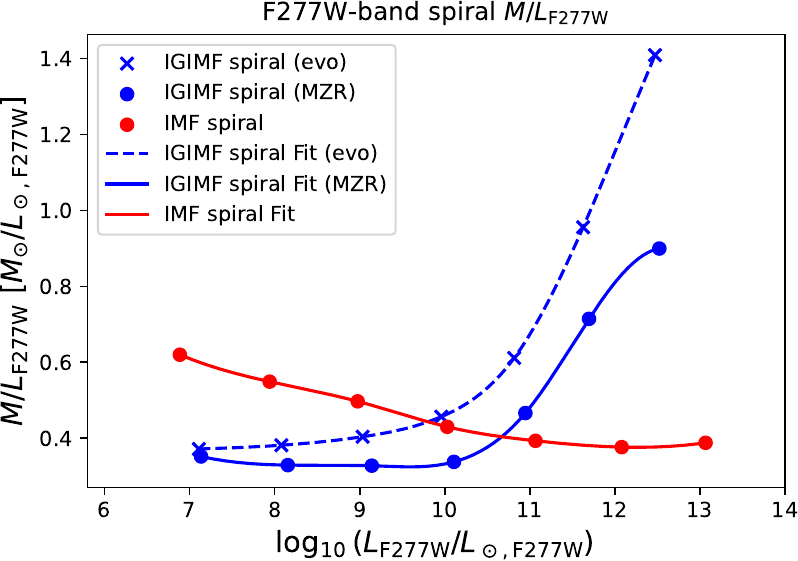}
    \caption{Same as Fig.~\ref{F090W}, but for the F277W band.}
    \label{F277W}
\end{figure}

In this section, we show the relation between $M_{\star}/L$ and luminosity for elliptical and spiral galaxies in the F115W, F150W, F200W, and F277W bands (Figures~\ref{F115W}, \ref{F150W}, \ref{F200W} and~\ref{F277W}, respectively).

\section{MOND strong lensing masses \lowercase{vs} IGIMF baryonic masses}
\label{lensing_vs_bar}

\begin{figure}
    \centering
    \includegraphics[width = 0.45\textwidth]{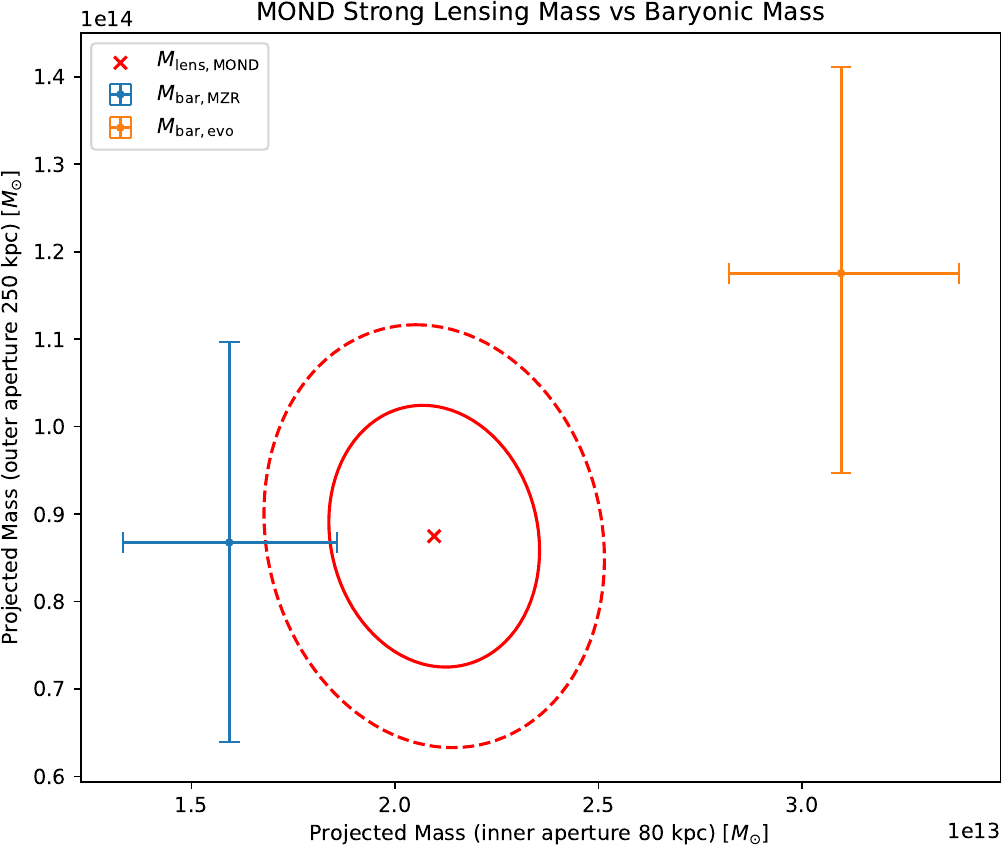}
    \includegraphics[width = 0.45\textwidth]{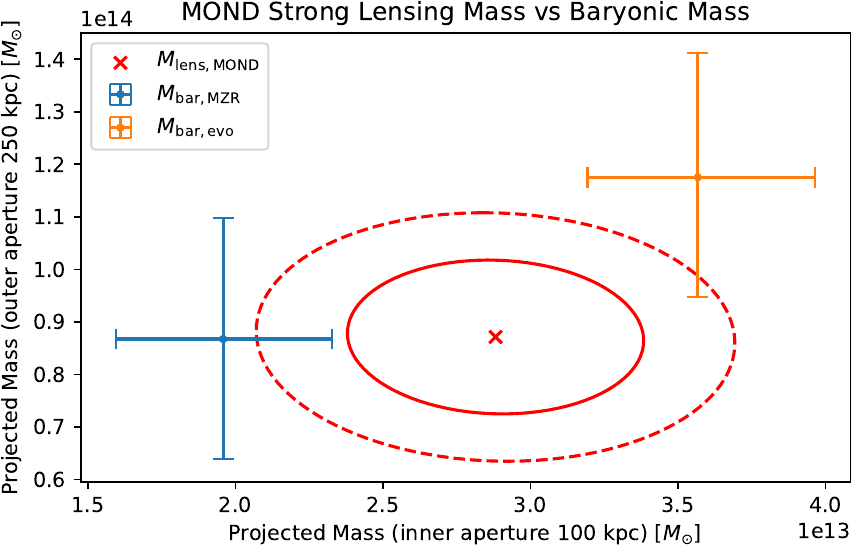}
    \includegraphics[width = 0.45\textwidth]{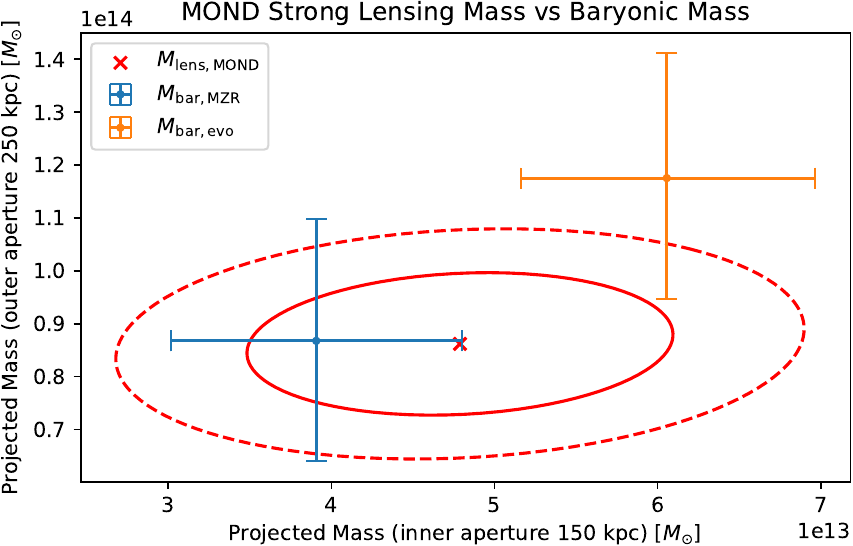}
    \caption{Comparison between the MOND strong lensing masses and the baryonic masses within different 2D projected radii of the southern core of the main clump. The $y$-axis represents the mass within 250~kpc, while the $x$-axis from top to bottom corresponds to the mass within 80, 100, and 150~kpc, respectively. The red crosses indicate the MOND strong lensing masses, while the solid and dashed red ellipses show the $1\sigma$ and $2\sigma$ uncertainty regions, respectively. The blue and orange dots with $1\sigma$ error bars denote the IGIMF baryonic masses estimated from the constant- (MZR) and enriched-metallicity (evo) models, respectively.}
    \label{kpc_vs_kpc}
\end{figure}

\begin{figure}
    \centering
    \includegraphics[width = \linewidth]{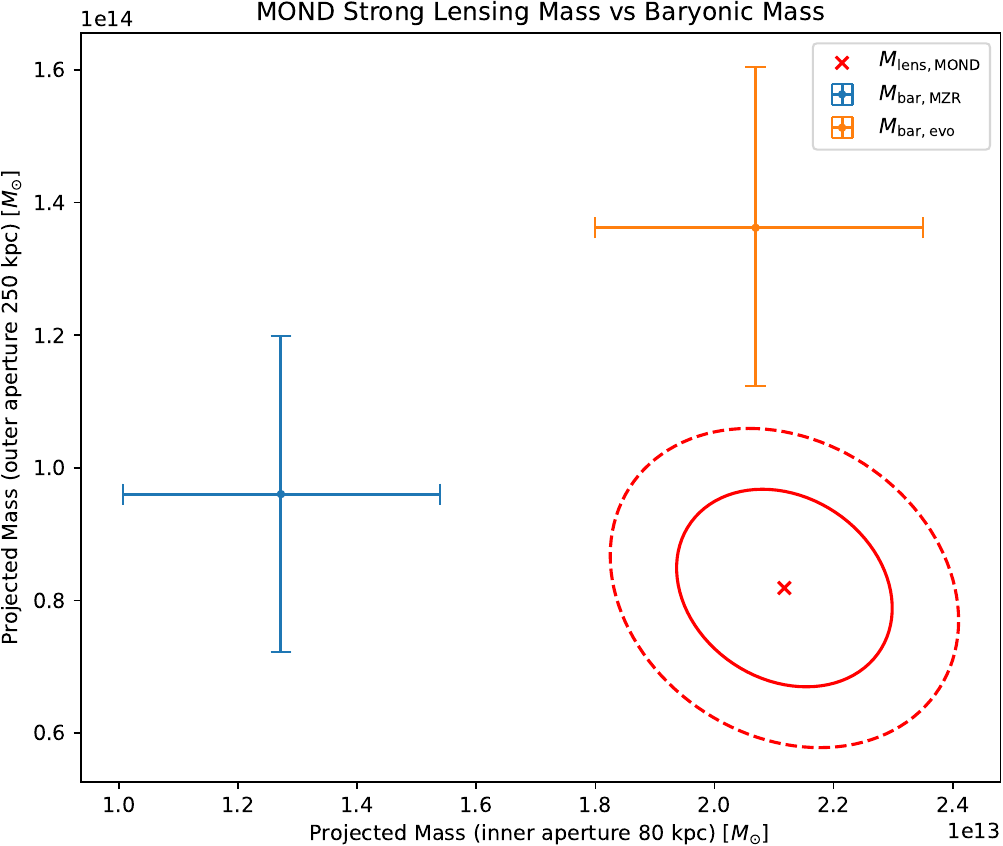}
    \includegraphics[width = \linewidth]{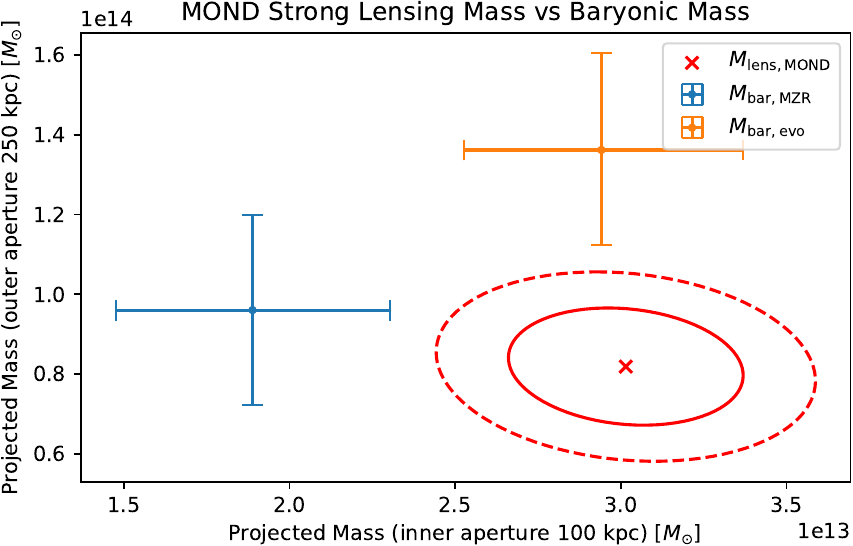}
    \includegraphics[width = \linewidth]{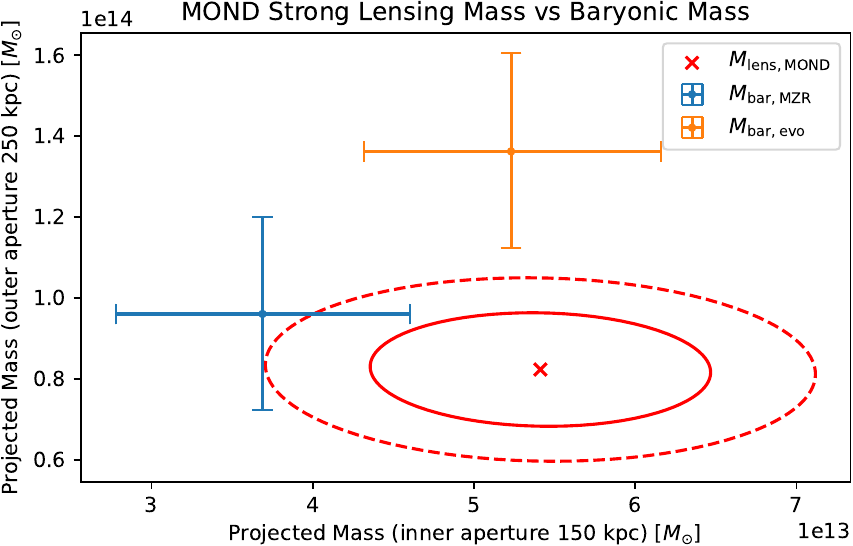}
    \caption{Same as Fig.~\ref{kpc_vs_kpc}, but for the northern core.}
    \label{kpc_vs_kpc_north}
\end{figure}

\begin{figure}
    \centering
    \includegraphics[width = \linewidth]{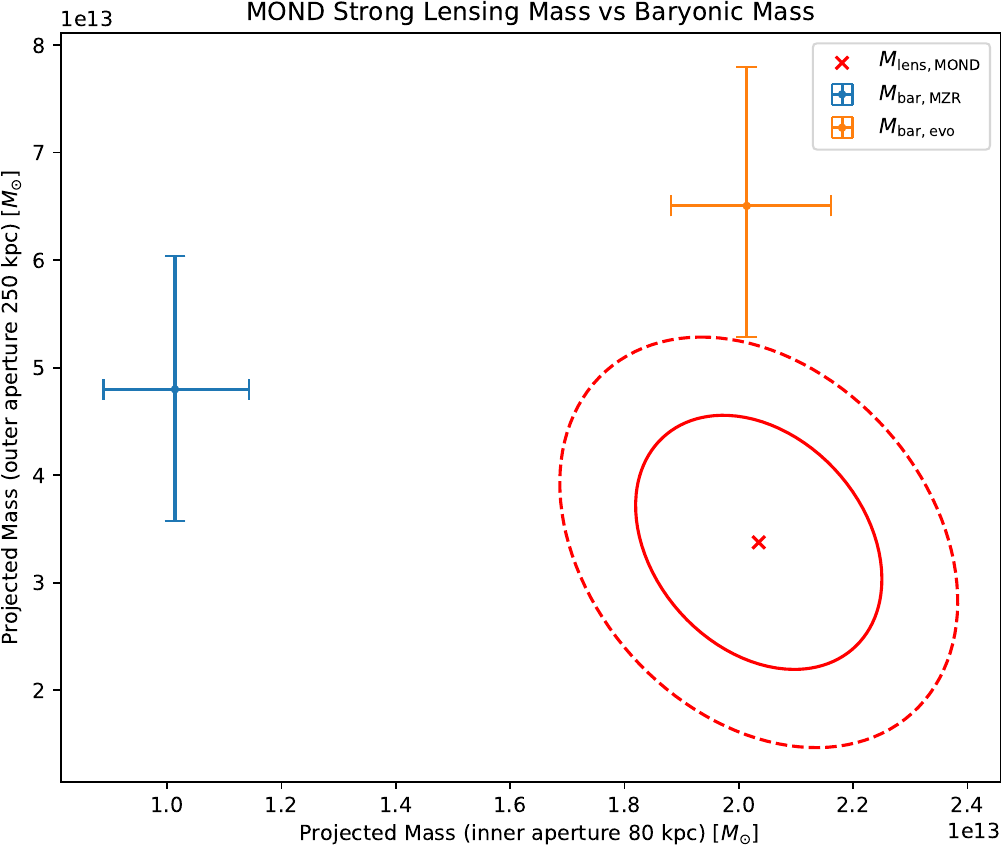}
    \includegraphics[width = \linewidth]{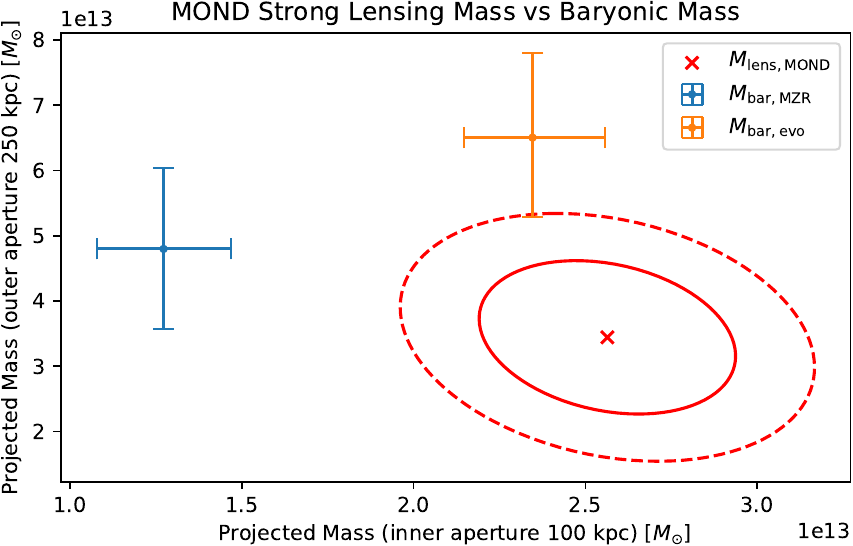}
    \includegraphics[width = \linewidth]{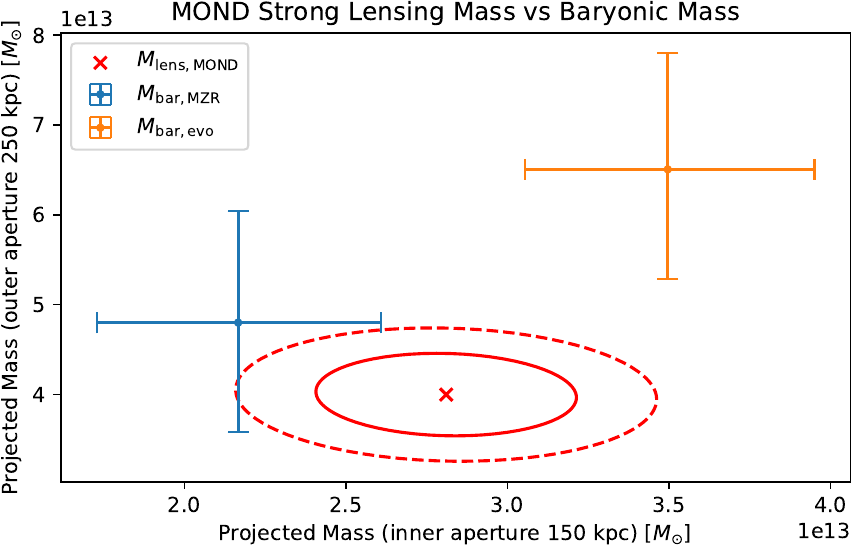}
    \caption{Same as Fig.~\ref{kpc_vs_kpc}, but for the subclump core.}
    \label{kpc_vs_kpc_sub}
\end{figure}

In Figs.~\ref{kpc_vs_kpc}, \ref{kpc_vs_kpc_north}, and \ref{kpc_vs_kpc_sub}, we present comparison plots of the total baryonic mass versus the MOND strong lensing mass for 80 vs 250~kpc, 100 vs 250~kpc, and 150 vs 250~kpc, corresponding to the southern, northern, and subclump cores, respectively. All MOND strong lensing masses are derived from the GR strong lensing results obtained by \citetalias{Cha_2025}. From these figures, it can be seen that the MOND strong lensing masses generally lie between the total baryonic masses predicted by the IGIMF models with constant and enriched metallicities. The enclosed MOND strong lensing masses within a 250~kpc aperture centred on all three cores are consistently closer to the masses obtained from the constant-metallicity model, possibly because this better describes the metallicity evolution of galaxies in lower density regions.

\end{appendix}

\label{lastpage}
\end{document}